\documentclass[english]{elsarticle}
\usepackage[T1]{fontenc}
\usepackage[latin9]{inputenc}
\usepackage{geometry}
\usepackage{babel}
\usepackage{array}
\usepackage{longtable}
\usepackage{float}
\usepackage{rotfloat}
\usepackage{booktabs}
\usepackage{url}
\usepackage{amssymb}
\usepackage{graphicx}
\usepackage[unicode=true]
 {hyperref}

\makeatletter

\newcommand*\LyXZeroWidthSpace{\hspace{0pt}}
\providecommand{\tabularnewline}{\\}
\floatstyle{ruled}
\newfloat{algorithm}{tbp}{loa}
\providecommand{\algorithmname}{Algorithm}
\floatname{algorithm}{\protect\algorithmname}

\usepackage{a4wide}
\usepackage{url}
\renewcommand{\citet}{\cite}
\usepackage{times}
\usepackage{algorithmicx}
\usepackage[noend]{algpseudocode}
\usepackage{mathtools}

\DeclarePairedDelimiter\abs{\lvert}{\rvert}
\algnewcommand\algorithmicforeach{\textbf{for each}}
\algdef{S}[FOR]{ForEach}[1]{\algorithmicforeach\ #1\ \algorithmicdo}

\newcommand{\xsca}[0]{X^{\mathrm{sca}}} 
\newcommand{\xsch}[0]{X^{\mathrm{plan}}} 
\newcommand{\itype}[0]{I^{\mathrm{type}}} 
\newcommand{\ischeme}[0]{I^{\mathrm{scheme}}} 
\newcommand{\tasks}[0]{T} 

\newcommand{\valueIteration}[0]{\textit{Value Iteration}} 
 
\newcommand{\qLearning}[0]{\textit{Q-learning}} 
\newcommand{\sarsa}[0]{\textit{SARSA}} 
\newcommand{\scalingH}[0]{Scaling (H)} 
\newcommand{\scalingV}[0]{Scaling (V)} 
\newcommand{\sched}[0]{Scheduling}
\newcommand{\wf}[0]{Workflow} 
\newcommand{\aCloud}[0]{Services} 
\newcommand{\tIndep}[0]{Independent Tasks}
\renewcommand{\textendash}{--}

\makeatother

\begin{document}
\begin{frontmatter}
\journal{Engineering Applications of Artificial Intelligence}
\title{Reinforcement Learning-based Application Autoscaling in the Cloud: A Survey}

\author[itic,conicet]{Yisel Gar\'i}
\ead{ygari@uncu.edu.ar}

\author[itic]{David A. Monge\corref{cor}}

\ead{dmonge@uncu.edu.ar}

\author[itic,fing,conicet]{Elina Pacini}
\ead{epacini@uncu.edu.ar}

\author[isistan,conicet]{Cristian Mateos}
\ead{cristian.mateos@isistan.unicen.edu.ar}

\author[itic,fing]{Carlos Garc\'ia Garino}
\ead{cgarcia@itu.uncu.edu.ar}

\cortext[cor]{Corresponding author.}

\address[itic]{ITIC - Universidad Nacional de Cuyo. Mendoza, Argentina}

\address[fing]{Facultad de Ingenier\'ia - Universidad Nacional de Cuyo. Mendoza, Argentina}

\address[conicet]{Consejo Nacional de Investigaciones Cient\'ificas y T\'ecnicas (CONICET). Argentina}

\address[isistan]{ISISTAN-UNICEN-CONICET. Tandil, Buenos Aires, Argentina}

\begin{abstract}
Reinforcement Learning (RL) has demonstrated a great potential for automatically solving decision-making problems in complex uncertain environments. RL proposes a computational approach that allows learning through interaction in an environment with stochastic behavior, where agents take actions to maximize some cumulative short-term and long-term rewards. Some of the most impressive results have been shown in Game Theory where agents exhibited superhuman performance in games like Go or Starcraft 2, which led to its gradual adoption in many other domains, including Cloud Computing. Therefore, RL appears as a promising approach for Autoscaling in Cloud since it is possible to learn transparent (with no human intervention), dynamic (no static plans), and adaptable (constantly updated) resource management policies to execute applications. These are three important distinctive aspects to consider in comparison with other widely used autoscaling policies that are defined in an ad-hoc way or statically computed as in solutions based on meta-heuristics. Autoscaling exploits the Cloud elasticity to optimize the execution of applications according to given optimization criteria, which demands to decide when and how to scale-up/down computational resources, and how to assign them to the upcoming processing workload. Such actions have to be taken considering that the Cloud is a dynamic and uncertain environment. Motivated by this, many works apply RL to the autoscaling problem in the Cloud. In this work, we survey exhaustively those proposals from major venues, and uniformly compare them based on a set of proposed taxonomies. We also discuss open problems and prospective research in the area.
\end{abstract}

\end{frontmatter}{}

\section{Introduction}

Cloud computing~\citet{Mauch2013} brings a technological solution
for the execution of different type of applications due to its reliability,
availability, and resources scalability. Particularly, the IaaS Cloud
service model allows users to create and destroy different types of
Virtual Machine (VM) instances, optionally under a pay-per-use scheme.
In this way, it is possible to dynamically adjust the infrastructure
according to the variations in resource demands during the execution
of applications. This feature of Clouds together with the need to
achieve efficient execution of applications encourages the study and
development of autoscaling strategies in the Cloud~\citet{Monge2017,Gari2019}.
These strategies are aimed to optimize the application execution based
on different objectives such as execution time and economic cost,
as well as compliance with restrictions or the Service Level Agreements
(SLA), if specified. An SLA represents an agreement between a service
provider and its users to define quality aspects of the service offered
by the provider based on user requirements~\citet{Kearney2011}.

Autoscaling strategies periodically solve two interrelated optimization
problems, i.e. scaling and scheduling. The scaling stage consists
of adjusting the number and type of Cloud resources acquired (e.g.,
VMs) according to the application demand. On the other hand, the scheduling
stage consists of assigning each application task to the acquired
resources. Both subproblems are NP-hard so they are usually approached
with heuristics. Considering that the variability in Cloud performance
represents an important factor of uncertainty in the execution of
applications, several recent investigations propose solutions based
on Reinforcement Learning (RL) to solve some of the involved subproblems,
the scaling stage~\citet{Dutreilh2011,Barrett2012,T.Veni2016,Arabnejad2017,Ghobaei-Arani2018,Dezhabad2018,BibalBenifa2018,Gari2019}
or the scheduling stage~\citet{Barrett2011,Peng2015,Xiao2017,Duggan2017,Liu2017,Soualhia2018,Cheng2018}.

RL is one of the three basic machine learning paradigms together with
supervised learning and unsupervised learning. Specifically, RL proposes
a computational approach that allows an agent to learn the appropriate
behavior to achieve its objective by interacting with a stochastic
environment~\citet{Sutton:2018}. The agent periodically takes an
action that modifies the state of the environment and observes a reward
signal that allows the agent to evaluate the immediate effect of the
action taken. Actions also have long-term consequences that are not
immediately perceptible. Therefore, RL purpose is to let the agent
learn appropriate policies \textendash i.e., mapping the states to
actions\textendash{} to generate the greatest long-term benefit following
the agent objective. RL-based strategies are being widely used and
very encouraging results have been obtained in areas such as Game
Theory~\citet{Silver2016,Mnih2015}, which has motivated its study
and application in other areas, concretely autoscaling in Clouds.

It is also important to highlight that the main motivation for addressing
the autoscaling of applications in Clouds from the perspective of
RL are the following:
\begin{enumerate}
\item Policies are \emph{transparent}, i.e., they are not dependent on human
intervention or deep domain knowledge, since the scaling and scheduling
policies are learned through interaction with the environment;
\item Policies are \emph{dynamic}, i.e., a learned policy determines the
most adequate action based on the current state of the environment
and the application execution, instead of a static plan previously
computed as in solutions based on meta-heuristics; and
\item Policies are \emph{adaptable}, i.e., \emph{online} policy learning
facilitates policy improvement and constant updates. Thus, learned
policies can adapt to the changes that occur in the dynamics of the
Cloud environment, unlike policies learned in \emph{offline} mode~\citet{Barrett2011,Gari2019}
that are prone to become obsolete in time.
\end{enumerate}
Therefore, in a Cloud setting \textendash i.e., an environment with
uncertainty\textendash{} where, for example, the variability in the
performance of VMs constantly changes, it is necessary to make appropriate
decisions on the fly. The use of RL enables feasible decision-making
regarding the type and number of VM to use (scaling) as well as which
resources should be assigned at any moment (scheduling). Motivated
by this, several recent researches propose RL-based approaches to
solve the Cloud autoscaling problem.

Hence, we have conducted a literature review of relevant works that
address the Cloud autoscaling problem via solutions based on RL, being
to the best of our knowledge the first survey on this topic. In particular,
we have surveyed Cloud autoscaling approaches for three types of applications,
namely workflows, independent tasks, and Cloud services. We have classified
the surveyed works based on a taxonomy according to the type of RL-based
technique used. From the analysis of the state of the art in the application
of RL strategies for autoscaling in Cloud, noticeable findings are
that none of the works jointly solves the scaling and scheduling subproblems.
Besides, although workflow is a mature technology driving many of
the Cloud applications nowadays, very few works have been proposed
for workflows and those that consider workflows only focus on scheduling
without taking into account the scaling problem. These facts altogether
evidence not only the promissory nature of RL-based application autoscaling
in Clouds but also the fertile characteristic of the area in terms
of prospective future improvements.

This article is organized as follows. In Section~\ref{sec:Background},
the background is presented, explaining underpinning concepts such
as Cloud application types, the Cloud Computing paradigm and related
provisioning models, the autoscaling problem, and the RL basics that
will be referred throughout work. Section~\ref{sec:RelatedWork}
discusses the relevant works in the area in detail. Then, in Section~\ref{subsec:Discussion}
the limitations of the surveyed works and open problems are highlighted.
Section~\ref{sec:Conclusions} concludes the survey. Finally, in
Appendix~A.1 through~A.4 we overview different relevant techniques
under the umbrella of RL that are exploited by the surveyed approaches.

\section{Background\label{sec:Background}}

In this section, the theoretical and technical foundations underpinning
the present paper are discussed, namely those related to Cloud applications,
Clouds as an execution environment, the concept of autoscaling, and
the RL basics. For this, the concepts and fundamental characteristics
of Cloud applications are presented in subsection~\ref{sec:Applications}.
Then, the Cloud Computing paradigm and its different service models
are analyzed in subsection~\ref{sec:CloudComputing}. Later, in subsection~\ref{sec:CloudAutoescaling}
we describe the Cloud autoscaling problem by integrating the two previous
topics. Finally, we introduce the RL basics necessary for a better
understanding of the present work.

\subsection{Cloud applications\label{sec:Applications}}

Applications in the Cloud can fall into one of three categories:
\begin{enumerate}
\item \emph{Workflows}. Broadly, a workflow describes a complex objective
through the composition of a set of tasks and their dependencies.
On the other hand, workflow technology focuses on the development
of applications that use readily-available software components. This
approach allows people without experience in programming languages,
\LyXZeroWidthSpace \LyXZeroWidthSpace but with solid knowledge of
the problem domain, to contribute to the development of new applications.
Workflows have been and are widely used in the modeling of complex
research experiments in several disciplines such as Geoscience~\citet{liu2016efficient},
Astronomy~\citet{Brown2007,meade2018evaluating}, and Bioinformatics~\citet{vandenbrouck2019bioinformatics},
among many others. Hence, workflows built in this context are termed
\emph{scientific workflows}.
\item \emph{Independent tasks}. Applications of this type are constructed
as a set of tasks logically related but without dependencies among
them. This kind of application is also known as \emph{bag of tasks}.
In this kind of application the tasks can be executed independently
without requiring the output of other tasks. Examples of these applications
are: different runs of a Monte Carlo simulation~\citet{Wang2011,Hu2019},
certain types of parameter sweep experiments~\citet{Monge2018},
and any type of embarrassingly parallel jobs. Note that independent-task
applications can be considered a particular case of workflow applications,
where all tasks depend on a single fictitious start task and precede
a single fictitious end task. And, just like in the case of workflows,
tasks are in general data-intensive and/or CPU-intensive, and hence
require a large amount of computational hardware and software resources
that include computing power, high-speed networks, storage capacity,
and sophisticated administration tools, among others.
\item \emph{Cloud services}. Applications of this type are the most dissimilar
to the previous ones. In general, they serve multiple users at a time
responding to \emph{user requests}. User requests (from now on tasks,
to homogenize nomenclature with the other two types of applications)
are independent of each other and can arrive unpredictably. Examples
of this are Facebook, Twitter, Google Drive, or any Cloud-hosted Web
application, just to name a few. Note that contrary to the previous
two kinds of applications, the number of tasks here can dramatically
vary over time and they usually have a short duration.
\end{enumerate}
A detailed explanation regarding the particularities of each application
type and its implications is given in Section~\ref{subsec:ApplicationTaxonomy}.
However, we close this section by highlighting that although every
type of application has its own particularities, all of them share
some characteristics. First, in all cases, there is a considerable
number of tasks that can be executed in parallel. Second, workloads
can heavily vary over time (e.g. either because of the workflow structure
itself or just because of the number of requests changes). Such variability
determines instants of time where an infrastructure with greater or
lesser capacity is required. The provisioning and elasticity capabilities
of the Cloud computing paradigm make it an excellent candidate to
meet the varying computational requirements of these applications.
The following section discusses the main features of this computing
paradigm.

\subsection{Cloud Computing\label{sec:CloudComputing}}

The National Institute of Standards and Technology (NIST) defines
Cloud Computing as ``a model for enabling ubiquitous, convenient,
on-demand network access to a shared pool of configurable computing
resources (e.g., networks, servers, storage, applications, and services)
that can be rapidly provisioned and released with minimal management
effort or service provider interaction~\citet{Mell2011}''.

The tendency to expose Everything as a Service (XaaS) when it comes
to Cloud capabilities describes a widely adopted scenario in which
service-oriented architecture and design principles underpin the development
and implementation of software services in Clouds~\citet{Duan2015,Fortino2014}.
In this sense, NIST defines the three following base service models
in the Cloud:
\begin{itemize}
\item Infrastructure as a Service (IaaS), where ``service'' means resource:
It is the most basic but at the same time ubiquitous model by which
an IT infrastructure is deployed in a data center as VMs. An IaaS
Cloud enables on-demand provisioning of computational resources in
the form of VM deployed in a datacenter, minimizing or even eliminating
associated capital costs for users, and letting those users adding
or removing capacity from their IT infrastructure to meet peak or
fluctuating resource demands. Examples of IaaS providers include Amazon
EC2\footnote{Amazon Elastic Compute Cloud: \url{https://aws.amazon.com/ec2}},
Windows Azure Services Platform\footnote{Windows Azure Services: \url{https://azure.microsoft.com}},
and Google Compute Engine\footnote{Google Cloud Platform: \url{https://cloud.google.com/}}.
\item Platform as a Service (PaaS), where ``service'' means platform-level
functionality: In this model, the user can create their own software
using tools and/or libraries from the provider, including operating
systems, programming languages, databases, and Web servers. Some examples
are Google App Engine and Windows Azure Cloud Services.
\item Software as a Service (SaaS), where ``service'' means application:
Under this model, providers install and operate application-level
software in the Cloud, which is accessed by users through the browser,
a mobile application or a Web Service API. Cloud users only interact
with installed Cloud applications through the Internet, but they do
not know where these applications are running or the implementation
and installation details. Examples of SaaS are Google Apps, Microsoft
Office 365, and Dropbox.
\item Function as a Service (FaaS), where ``service'' is referred to as
a \textquotedblleft runtime software component in a serverless architecture\textquotedblright :
It is based on lightweight functions that can be triggered by a given
event. Building an application following this model means just writing
functions without pondering about concerns such as deployment, server
resources, scalability, etc. The most prominent example is AWS Lambda,
but there are other alternatives such as Google Cloud Functions, Microsoft
Azure Functions, and Webtask.io.
\end{itemize}
This work is placed in the context of the service model IaaS offered
by public Cloud providers because the surveyed works mainly focus
on how the virtual infrastructure is scaled when running resource-intensive
Cloud applications.

One of the main features of the IaaS model is the elasticity at the
infrastructure level, which allows users to dynamically acquire and
adjust the computing infrastructure according to their needs. The
elasticity in IaaS is supported from a technical perspective through
the use of virtualization technologies~\citet{Buyya2009}. Virtualization
technologies allow us to share the resources of a single physical
machine (PM) among several independent VM instances. Several VMs might
co-exist in the same PM and have no visibility or control over the
configuration of the PM that hosts them or over the neighboring VMs.
Depending on the configuration, each VM has assigned a portion of
the physical resources available in the PM (CPU, RAM, storage, and
network bandwidth). Then, a VM monitor installed in the PM is responsible
for controlling the access of each VM to the physical resources. The
VM monitor tries to isolate individual VMs from its environment for
security reasons and possible failures, but not to improve performance~\citet{Koh2007,Pu2010}.
Consequently, the unpredictable performance of the VMs~\citet{Koh2007,Pu2010}
becomes one of the main obstacles facing the Cloud computing model~\citet{Armbrust2009}.

Another key feature of the IaaS model is the eradication of costs
associated with the maintenance of the infrastructure since users
only pay for the resources they use. In this way, users are granted
access to various types of VMs under a pay-per-use scheme, with a
wide range of hardware and software configurations. Typically, prices
differ according to the computing capabilities of available VMs, but
prices may also differ depending on the \emph{price model} under which
a VM is rented. Two of the most common pricing models are:

\paragraph{On-demand/Non-preemptible instances}

This price model is suitable for users with sporadic and bursting
demands since the on-demand option allows users to rent resources
right away without a fixed time limit, as opposed to \emph{reserved}
instances, where the user rents resources for a fixed time duration
while obtaining important price discounts. However, on-demand instances
generally have higher prices than reserved instances, considering
the same computing capabilities for acquired VMs. Besides, the on-demand
option adopts billing cycles generally per hour of use, so partial
use of the instances is rounded up to one hour\footnote{Amazon recently made adjustments to its pricing policies by switching
to a billing variant per second, in this way the user pays only for
the time used \href{https://aws.amazon.com/es/ec2/pricing/}{https://aws.amazon.com/es/ec2/pricing/}}.

\paragraph{Spot/Preemptible instances}

Cloud providers have unsold computing capacity during certain periods.
To encourage users to buy additional capacity, they offer the option
of spot instances. The prices of spot instances fluctuate over time,
but they are usually much lower (up to 90\% in some cases) than the
prices of on-demand instances, considering the same computing capabilities.
Then, the user makes an offer with the maximum value that is willing
to pay for the instance and will be charged only with the (much cheaper)
spot price at any time. When the spot price exceeds the user's bid,
the VM instances are terminated interrupting any running task on them.

Some IaaS providers such as Amazon have recently decided to no longer
rely on the fluctuation of VM prices based on bid schemes~\citet{Fabra2019}.
In the new model, the spot prices are more predictable, updated less
frequently, and are determined by supply and demand for Amazon EC2
spare capacity, not bid prices. However, factors that condition the
interruption of a spot instance are completely internal to the provider
and cannot be known without a deep understanding and analysis of the
AWS infrastructure at runtime. Then, it is important to consider,
together with the economic advantages of spot instances, the fact
that spot instances are less reliable since the tasks they execute
are subject to sudden, unpredictable terminations. User-provided checkpoints
mechanisms are necessary to avoid losing task progress when the instance
in which the task is running is terminated by the infrastructure.

For simplicity, from now on we will refer to the different instances
types according to the terminology used by Amazon (reserved, on-demand,
and spot). These pricing models offer a wide range of options to shape
the infrastructure they need for the execution of their applications,
and the elasticity that enables the dynamic reconfiguration of such
infrastructure.

\subsection{Autoscaling in Clouds\label{sec:CloudAutoescaling}}

As stated at the end of Section \ref{sec:Applications}, the workload
associated with Cloud applications presents a lot of variabilities
over time. Such characteristics stress the necessity for resource
elasticity. On top of that, these applications can be compute-\-intensive
and/or require the processing of large volumes of data (e.g. workflows
and independent tasks), or simply need to execute a massive number
of lightweight tasks that, when aggregated, leading to high demand
for resources (e.g. services). All these applications usually involve
hundreds, thousands, or even a larger number of tasks with varying
durations that can range from a few minutes to several days or weeks,
optionally processing from MiB to TiB of input data. This adds a new
complexity dimension as the duration of the tasks may further differ
depending on the type of VM acquired to run them. Besides, since tasks
usually have different execution profiles (processing load and memory
usage), some types of instances may be more suitable for certain application
types. For example, Amazon offers different instance types to be used
according to the type of problem to solve. Some of them are general-purpose
instances, which provide a balance of computing, memory, and networking
resources, and are suitable for applications that use these resources
in equal proportions such as web servers and code repositories. Moreover,
compute-optimized instances are ideal for compute-bound applications
that benefit from high-performance processors, for example, media
trans-coding, high-performance computing (HPC), scientific modeling,
dedicated gaming servers, and machine learning inference.

Even more, in public Clouds, the use of resources has an economic
cost, and the different types of instances differ in their price.
When having in mind maximizing the execution efficiency (e.g. reducing
the execution cost, time, etc.) acquiring the appropriate infrastructure
(i.e. the number of VMs of each type and their price models) becomes
a complex problem.

All in all, autoscaling strategies have to deal with the dynamic scaling
of the infrastructure according to the application needs and the on-line
scheduling of such tasks on the running infrastructure. These are
two interdependent problems that must be solved in tandem~\citet{Monge2017}:
\begin{enumerate}
\item \emph{Scaling}, which consists of dynamically adjusting (i.e provisioning
or releasing) the available virtualized resources for more efficient
use of them. Scaling~\citet{Krzywda2018} can take two forms: horizontal
scaling, when the number of assigned VMs of any type to an application
can dynamically vary through its execution, and vertical scaling,
when the capabilities of individual VMs (CPU, memory, I/O) are varied
without hindering the execution of the applications in such VMs. In
the second case, this is usually achieved by replacing one VM with
another VM having different capabilities, and performing task migration.
\item \emph{Scheduling}, which consists of assigning tasks for execution
in the acquired VM instances.
\end{enumerate}
Both subproblems are NP-hard and, therefore, the solutions proposed
to date are mainly based on heuristics~\citet{Mao2013,Monge2017,Monge2018,Monge2020}.

\begin{figure}[H]
\begin{centering}
\includegraphics[clip,width=1\columnwidth]{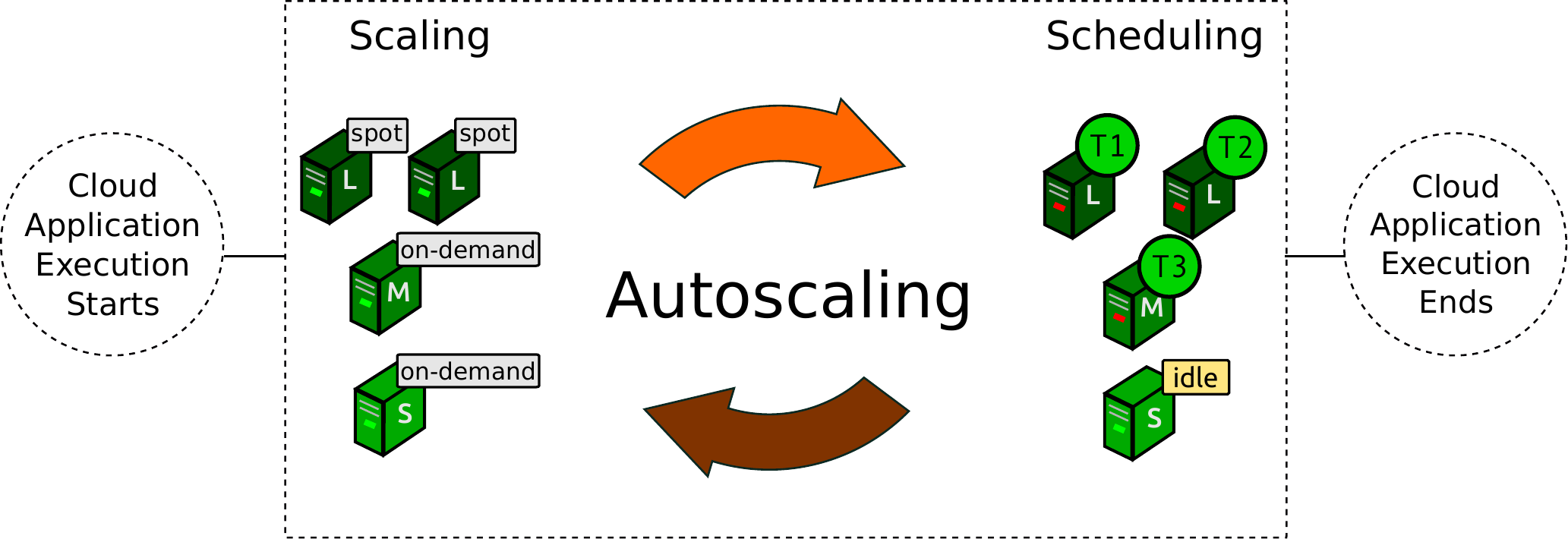}
\par\end{centering}
\caption{\label{fig:Autoescalado}Cyclic autoscaling process that includes
the scaling and scheduling subprocesses. $L$, $M$ and $S$ represent
different VM capabilities. $T_{i}$ represent tasks to execute.}
\end{figure}

In a Cloud infrastructure, the demand for resources fluctuates over
time, and the VMs located in the same PM constantly compete for the
resources they share. Also, the potential variations in network traffic
impact the communication speed between physically separated VMs. All
these elements make the performance of the Cloud infrastructure to
vary, which represents an important uncertainty factor in the decision-making
processes behind scaling and scheduling since the real duration of
tasks usually differs from the estimated values. Then, autoscaling
strategies need to frequently update the information they have available
for taking those decisions, considering that:
\begin{itemize}
\item applications present variable workload patterns at different stages,
\item models that estimate task durations are imperfect, and
\item as said, Cloud infrastructures are characterized by variable performance.
\end{itemize}
Then, autoscaling strategies need to monitor the state of the environment
(infrastructure and applications) to mitigate the effects of discrepancies
between the available information about tasks (estimations) and the
real progress of the execution, and hence making the most appropriate
scaling and scheduling decisions at runtime. In this sense, autoscaling
strategies are executed periodically as shown in Figure~\ref{fig:Autoescalado}.
In each update interval, autoscaling strategies adjust the number
of instances for each VM type and price model and assign the tasks
to the currently available VMs. 

In each update interval, the autoscaling strategy addresses an optimization
problem based on certain objectives of interest such as load balancing,
throughput, flowtime, energy consumption, makespan, monetary cost,
and so on. It is important to mention that both the makespan and cost,
or a combination of both, are the optimization objectives most addressed
by researchers in the context of Cloud scheduling~\citet{Pacini2014}.
More formally, given the set $\itype$ of instance types (in terms
of hardware and software capabilities) considered for autoscaling,
the set $\ischeme$ of price models (reserved, spot, and on-demand
instances), the set $\tasks$ of tasks to process, and the set $I$
of available instances in the current infrastructure, then, in each
update interval the autoscaling strategies generate:
\begin{itemize}
\item $\xsca=\{\itype\times\ischeme\rightarrow\mathbb{N}_{0}\}$, a \emph{scaling}
plan that indicates the required number of reserved, on-demand, and
spot instances of each type, and
\item $\xsch=\{\tasks\rightarrow I\}$, a set of \emph{scheduling} decisions
that map each task $t\in\tasks$ to one of the $i\in I$ instances.
\end{itemize}
The problem of optimizing application execution in the Cloud with
autoscaling strategies can be addressed from different perspectives:
the execution of individual applications from different users, the
execution of multiple applications from the same user, or many applications
from different users. In all cases, it is important to consider that
when exploiting non-dedicated IaaS Clouds, the performance can be
affected by workloads external to the applications being executed
themselves.

Thus, although most of the proposed solutions for the autoscaling
problem are based on heuristics or meta-heuristics, recent researches
aim to apply reinforcement-learning approaches to solve any of the
subproblems involved, i.e. the scaling~\citet{Dutreilh2011,Barrett2012,T.Veni2016,Arabnejad2017,Ghobaei-Arani2018,Dezhabad2018,BibalBenifa2018,Gari2019}
or scheduling~\citet{Barrett2011,Peng2015,Xiao2017,Duggan2017,Liu2017,Soualhia2018,Cheng2018}.

\subsection{Reinforcement Learning}

Reinforcement learning (RL)~\citet{Sutton:2018} is one of three
basic machine learning paradigms, alongside supervised learning and
unsupervised learning. RL is concerned with how a software agent ought
to take actions in an environment to maximize some notion of cumulative
reward. In other words, the computational approach where an agent,
acting in an environment with uncertainty, learns to associate situations
with actions while maximizing a numerical reward signal is considered
reinforcement learning. At the beginning, the agent does not know
what actions to take, and as time passes, it must discover which actions
produce the greatest long-term benefit, trying again and again, different
options. In the most interesting and challenging cases, actions may
affect not only the immediate reward but also the next situation and,
through that, all subsequent rewards. Making the most appropriate
decision requires taking into account the indirect consequences of
the actions, therefore, some kind of foresight or planning is necessary.
These two characteristics, \emph{trial-and-error search,} and \emph{delayed
reward}, are the two most important distinguishing features of RL~\citet{Sutton:2018}.

\begin{figure}[H]
\begin{centering}
\includegraphics[scale=0.8]{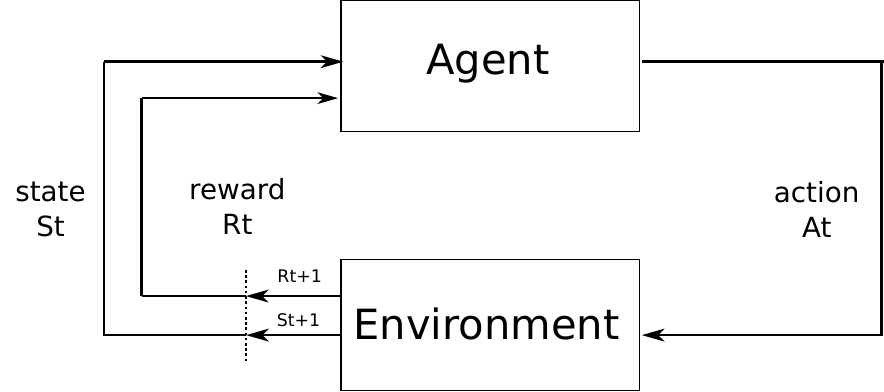}
\par\end{centering}
\caption{\label{fig:agent-environment}MDP interaction process between an agent
and the environment (Figure adapted from~\citet{Sutton:2018}).}
\end{figure}

Markov decision processes (MDP)~\citet{Bellman1957} provide a formal
framework widely used in the context of RL to define the interaction
between a learning agent and its environment in terms of states, actions,
and rewards (see Figure~\ref{fig:agent-environment}). MDPs have
become the de facto standard formalism for learning sequential decision-making~\citet{Otterlo2009}
and it has been applied to autoscaling problems in Cloud~\citet{Tang2012,Barrett2011,Barrett2012}.
The classical model of an MDP is defined as a 5-tuple $(S,A,P_{\cdot}(\cdot,\cdot),R_{\cdot}(\cdot,\cdot),\gamma)$,
where :
\begin{itemize}
\item $S$ represents the environmental state space;
\item $A$ represents the whole action space;
\item $P_{a}(s,s')=Pr(s_{t+1}=s'\mid s_{t}=s,a_{t}=a)$ represents the probability
that action $a$ in state $s$ at time $t$ will lead to state $s'$
at time $t+1$;
\item $R_{a}(s,s')$ represents the (expected) immediate reward received
after transitioning from state $s$ to state $s'$ due to action $a$;
\item $\gamma\in[0,1]$ (or discount factor) is the difference in importance
between future and immediate rewards. When $\gamma$ is close to 0,
rewards in the distant future are viewed as insignificant. When $\gamma$
is 1 all rewards are equally important.
\end{itemize}
As shown in Figure~\ref{fig:agent-environment}, the agent and the
environment continuously interact in a constant exchange process.
At each time $t$, the agent receives a representation of the environment
state $s_{t}\in S$, and then selects an action $a_{t}\in A$. In
the next step, and as a consequence of the executed action, the agent
receives a numerical reward signal $r_{t+1}\in R\subset\mathbb{R}$
and goes to a new state $s_{t+1}$. The boundaries between the agent
and the environment are determined by everything that the agent may
or may not control, and not by those things the agent knows or does
not know. For example, the agent can have knowledge of the current
status or how the reward is computed, but only has control over the
actions it takes. The reward is a numerical signal and the agent goal
is to maximize the total amount of received reward. In general, the
agent attempts to maximize the expected gain $G_{t}$ that is defined
as a specific function of the reward sequence. In the simplest case,
the gain is the sum of the rewards: $G_{t}=r_{t+1}+r_{t+2}+...+r_{T}$
where $T$ is a final time step. Then, considering the discount factor
$\gamma$, which determines the degree of importance of future rewards
compared to the immediate rewards, the agent will attempt to select
the action $a_{t}$ so that the sum of the discounted rewards is maximized.
In this case, the expected gain is calculated as: $G_{t}=r_{t+1}+\gamma r_{t+2}+\gamma^{2}r_{t+3}...=\sum_{k=0}^{\infty}\gamma^{k}r_{t+k+1}$.

Moreover, there are four fundamental elements in the RL learning process:
the policy, the reward signal, the value function, and optionally,
the model of the environment:
\begin{enumerate}
\item A \emph{policy} defines how the learning agent behaves at a given
time. A policy $\Pi:S\longrightarrow A$ is a mapping from perceived
states of the environment to actions to be taken when in those states.
\item A \emph{reward signal} evaluates the immediate effect of the taken
actions considering the goal of the RL problem faced. The reward is
a scalar signal that defines what are the good and bad events for
the agent and it is the primary basis for altering the policy.
\item A \emph{value function} specifies what is good in the long term and
it is fundamental for policy improvement. The state-value function
$V(s)$ is the expected gain (i.e cumulative reward over the future)
to be obtained starting from this state. Value functions predict rewards
into the future following a specific policy and the purpose of estimating
values is obtaining an improved policy to achieve more reward. Rewards
are given directly by the environment, but values must be estimated
and re-estimated from the sequences of observations an agent makes
over its entire lifetime. In some cases, the state value function
is not sufficient in suggesting a policy and it is required to estimate
the values related to each action. The action-value function $Q(s,a)$
represents the expected gain considering the state-action pair.
\item The \emph{model of the environment} mimics the dynamic that determines
how the environment behaves. Given a state and an action, the model
might predict the next state and the next reward. Models are used
for deciding on a course of action by considering possible future
situations before they are experienced (in \emph{offline} mode). However,
in many interesting problems, an accurate model of the environment
is not always available or the dynamic of the environment is prone
to change over time. In these cases, learning is based on current
situations being experienced (in \emph{online} mode). Methods for
solving RL problems that use models are called \emph{model-based}
methods, as opposed to simpler \emph{model-free} methods that are
explicitly trial and error learners.
\end{enumerate}
When solving an MDP \textendash i.e. obtaining an appropriate policy\textendash{}
two fundamental processes are present in the existing resolution strategies:
\begin{itemize}
\item The process of \emph{predicting }the policy, where the values of the
states or state-action pairs are estimated (i.e the function $V(s)$
or $Q(s,a)$ is updated), generally based on the current estimated
policy and the information of the environment (either from the model
or from experience, according to the technique).
\item The process of \emph{control} or improvement of the estimated policy,
where $\Pi(s)$ is computed, based on the current estimated values.
\end{itemize}
Both processes determine a continuous interaction between the different
approximations of the value function and the policy. This interaction
in time converges to the optimal values for both functions~\citet{Sutton:2018}.
Two of the most commonly used methods to solve MDPs are Dynamic Programming
methods (DP, see Appendix~A.1) and Temporal Difference methods (TD,
see Appendix~A.2). In the context of Autoscaling in Cloud, some proposals
also combine RL with Neural Networks (see Appendix~A.3) and Fuzzy
Logic (see Appendix~A.4). Next section discusses the current state
of the art regarding autoscaling in public Clouds by using RL techniques.

\section{Review of Cloud Autoscaling based on RL Techniques \label{sec:RelatedWork}}

The two subproblems of Cloud autoscaling \textendash i.e. scaling
and scheduling\textendash{} have been particularly addressed in the
literature as decision-making problems in stochastic environments.
The actions related to scaling might consist e.g. of increasing or
reducing the number of VMs in the virtual infrastructure, while the
actions related to scheduling consist of assigning each task to a
specific acquired VM. Due to the uncertainty in these subproblems,
proposals that model the autoscaling problem as an MDP have appeared,
and essentially they use different RL techniques to learn adequate
scaling~\citet{Dutreilh2011,Barrett2012,T.Veni2016,Arabnejad2017,Ghobaei-Arani2018,Dezhabad2018,BibalBenifa2018,Gari2019}
or scheduling~\citet{Barrett2011,Peng2015,Xiao2017,Duggan2017,Liu2017,Soualhia2018,Cheng2018}
policies. These policies allow an autoscaler to determine which action
is more convenient at any time to optimize a long-term objective.

As we describe in detail in Appendix~A.1 through~A.4, there are
different techniques for obtaining adequate policies. On the one hand,
\emph{Model-based} techniques require a perfect model of the environment
to compute an appropriate policy in \emph{offline} mode. The surveyed
proposals which fall in the Model-based category use \emph{Value Iteration}
(see sub-appendix~A.1), a well-\-known algorithm based on Dynamic
Programming. On the other hand, there are the so-called \emph{Model-free}
techniques, which allow an agent to obtain a proper policy in \emph{online}
mode without requiring a perfect model of the environment. In other
words, the policy is learned and improved over time in a process of
continuous interaction with the environment. The surveyed papers found
in the Model-free category use \emph{Q-learning} and \emph{SARSA}
(see sub-appendix~A.2), two reference algorithms for Temporal Difference
learning. As we mentioned earlier, RL techniques are usually affected
by large state spaces, which directly impacts the performance of the
aforementioned algorithms in terms of the time to compute a solution
and memory usage. In this sense, the use of non-linear functions to
approximate $Q(s,a)$ has been proposed, and solutions that combine
RL with deep neural networks, i.e. Deep Reinforcement Learning (DRL)
(see sub-appendix~A.3) have appeared. Some proposals use Fuzzy Logic
(FL) to represent rules capable of \emph{fuzzily} encompassing multiple
states in the context of RL, i.e. Fuzzy Reinforcement Learning (FRL)
(see sub-appendix~A.4).

\begin{figure}[H]
\begin{centering}
\includegraphics[scale=0.6]{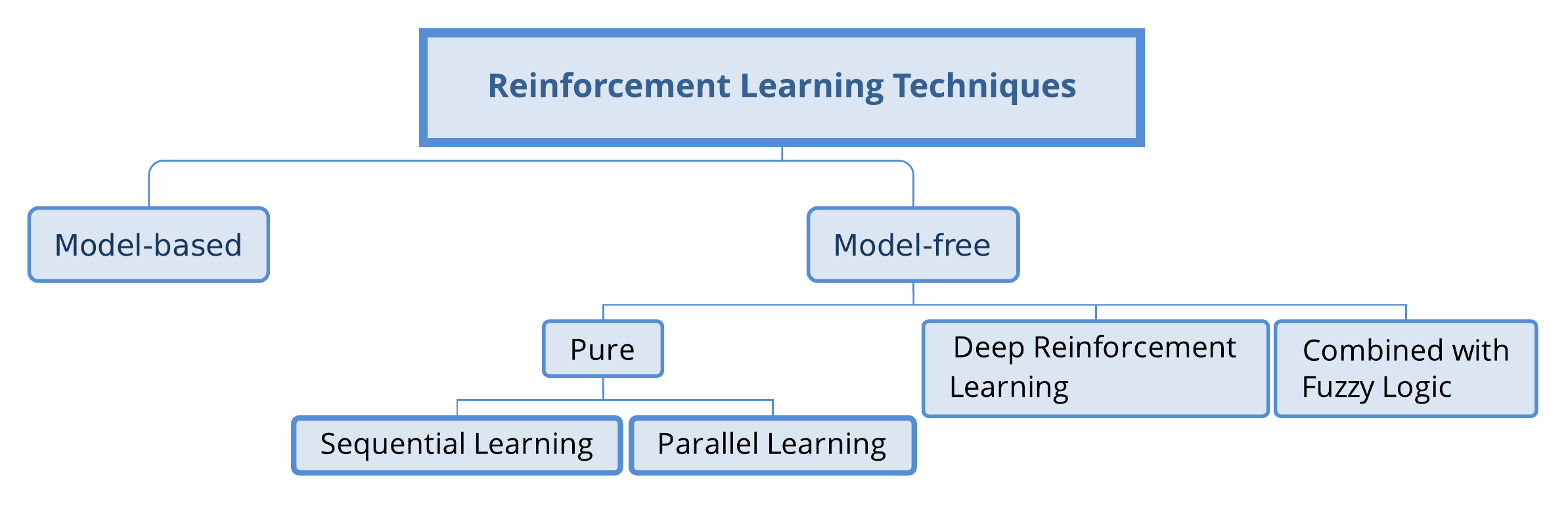}
\par\end{centering}
\caption{\label{fig:TaxoTechniques}Classification of RL based techniques applied
to the Cloud autoscaling problem.}
\end{figure}

In the next subsections, relevant works addressing the Cloud autoscaling
problem via solutions based on RL are described and analyzed. The
works are first organized according to the type of technique used
as defined in the taxonomy depicted in Figure~\ref{fig:TaxoTechniques}.
On the first level of the taxonomy, proposals in Model-based and Model-free
categories are presented. Then, on a second level, proposals on the
Model-free category are classified into three groups. First, are those
proposals that apply the technique in its original or pure formulation.
These techniques are further subdivided into sequential or parallel
since the variant of RL given by (multi-thread or multi-process) parallel
learning is distinctive. Second, we present the proposals that combine
RL with neural networks, and finally, the proposals that combine RL
with FL.

To perform the literature selection process, we focused our search
on the main citation databases (Google Scholar, Scopus, ACM Digital
Library, IEEE Explore). We generated a corpus of papers whose abstract
and title matched combinations of the following keywords: ``reinforcement
learning'', ``markov decision process'', ``autoscaling'', ``scheduling'',
``scaling'', ``workflow'', ``scientific application'', ``parallel
computing'', ``distributed computing''. Articles were first filtered
based on relevance considering the purpose of this survey. Then, a
second selection process was carried out based on the quality of the
journal/conference in which it was published. We kept articles published
in journals in the first two quartiles (Q1 and Q2) of the SCImago
Journal Rank, international conference papers (category A according
to ERA ranking and categories A{*}/B{*} according to the Qualis ranking)
and few papers not included in the aforementioned categories but very
relevant to the scope of this work. The result was a corpus of 21
articles.

Furthermore, to better structure the discussion of the selected works,
the following characteristics have been taken into account when describing
each work:
\begin{itemize}
\item The type of problem that is solved (i.e., scaling, scheduling, or
both if applicable),
\item The targeted optimization objectives,
\item The context of the problem (for example migrations of VMs, scheduling
of firewalls, etc.),
\item The considered variables in the definition of the states and reward
function,
\item RL-based algorithm used,
\item Baselines used for comparisons and experimental results (if reported),
\item Identified limitations.
\end{itemize}
Next, each of the works is classified by the type of RL technique
implemented (Model-based approaches in subsection~\ref{subsec:approachesMB}
and Model-free approaches in subsection~\ref{subsec:approachesMF}),
and then, in subsection~\ref{subsec:AnalysisState-of-the-art}, three
taxonomies based on the considered characteristics in the surveyed
works are proposed.

\subsection{Model-based Approaches \label{subsec:approachesMB}}

There are only two proposals~\citet{Barrett2011,Gari2019} in the
Model-based category. Both proposals have in common, on the one hand,
the estimation of the probability distribution of the transition between
states, due to the requirement of having a complete model of the environment
in Model-based methods. On the other hand, both works share the limitation
of learning the policies in an offline mode while operating in a dynamic
environment (Cloud infrastructure).

Barrett et al.~\citet{Barrett2011} propose an approach for the efficient
scheduling of workflows in the Cloud to minimize the makespan and
monetary cost under deadline constraints. First, a genetic algorithm
(GA) allows the approach to evolve different execution plans, where
each workflow task is assigned to one of the available VMs. Then,
through an MDP formulation and \emph{Value Iteration,} a policy that
dynamically chooses among the evolved plans the most suitable one
for the moment is obtained. For the states definition, the authors
consider three variables: a timestamps within the workload period
(0-23 hour), the resource load (light, moderate or heavy), and the
execution result (workflow failed or executed successfully). The actions
represent the selection of a specific scheduling plan. Then, the number
of actions depends on the number of GA solvers. The reward is computed
considering the cost of the workflow execution and a penalty if the
schedule results in a violation of the specified deadline. In~\citet{Barrett2011},
since there is no perfect model of the environment, the probability
distribution of the transitions between states is estimated from the
information obtained from multiple previous workflow executions. In
this sense, there is a limitation by which the quality of the obtained
policy will depend directly on the quality of the estimate of $P(s,a)$.

Also, Gar\'i et al.~\citet{Gari2018,Gari2019} study the learning of
budget allocation policies for the autoscaling of workflows in Clouds.
In such works, through the outputs of multiple workflow executions,
an MDP model is built, and then through the use of \emph{Value Iteration}
appropriate policies are derived. The derived policies are instead
used by a workflow autoscaling strategy called SIAA~\citet{Monge2017}
to determine in each autoscaling cycle, the adequate proportion of
spot versus on-demand instances that must be maintained. For the states,
the authors considered two features related to workload (i.e. the
proportion of long duration tasks and the maximum parallel degree
for the next period), the current budget limit, and the probability
of out-of-bid error (i.e. failures of the acquired spot instances
due to low bid). The values of each feature were discretized resulting
in 192 possible states. The actions represent the possible budget
assignment ratio between spot and on-demand instances (11 possible
values). The reward is computed as the ratio between the progress
and the cost of the workflow execution in the last cycle. Both in~\citet{Gari2019}
and~\citet{Barrett2011} there is a limitation given by the quality
of the obtained policy depends on the quality of the estimate for
$P(s,a)$. Besides, the fact that the policy is learned in an \emph{offline}
mode, gives the autoscaler partial ability to smoothly adapt to changes
in the environment at runtime. For example, the prices of the spot
instances and/or the probability of their failures could be subject
to variations. Therefore, if the policy was learned in online mode,
it could incorporate the experience of new executions and better adapt
to changes accordingly.

\subsection{Model-free Approaches \label{subsec:approachesMF}}

Unlike Model-based methods, Model-free methods adopt an \emph{online}
learning strategy and do not require a perfect model of the environment.
In this group is the largest number of related works in the area of
this survey and we will categorize them according to the classification
depicted in Figure~\ref{fig:TaxoTechniques}: pure proposals with
sequential learning (subsection~\ref{subsec:sequentialProposals}),
pure proposals with parallel learning (subsection~\ref{subsec:parallelProposals}),
proposals combined with neural networks (subsection~\ref{subsec:combinedNN})
and proposals combined with fuzzy logic (subsection~\ref{subsec:combinedFL}).

\subsubsection{Pure Proposals with Sequential Learning\label{subsec:sequentialProposals}}

In this section, we describe the works that use Model-free techniques
in their original formulation, as opposed to the proposals based on
DRL or FRL, and with a sequential learning process. At each decision
time, the value of a \emph{single} state-action pair is updated in
the table of values $Q(s,a)$. In this sense, these proposals are
more likely to have long training times since the speed of convergence
of the RL algorithms depends directly on the dimension of the state
space and actions.

Peng et al.~\citet{Peng2015} propose an approach to optimize task
scheduling in Cloud. The proposal is based on RL and queuing theory.
The states reflect the remainder of the buffer capacity of each VM.
The authors use a state aggregation technique to accelerate the learning
process with \emph{Q-learning}. Thus, the capacity of a VM is divided
into 5 possible categories (i.e. full, less, middling, more, vain).
The actions represent the selection of one of the available VMs for
scheduling the current task. The reward is 1, 0, or -1, considering
the mean waiting time of recent user requests and if the VM selected
was the one with the maximum capacity. Then, for experimentation,
two types of methods for task submission were defined: \emph{individual}
and \emph{grouped scheduling} methods. In \emph{individual scheduling},
user requests arrive continuously (as in regular Web requests or database
queries) and an immediate response is required. In this context, the
proposal outperformed in terms of response time to strategies such
as FIFO, \emph{fair-scheduling}~\citet{Jang2019}, \emph{greedy-scheduling}~\citet{Dong2015},
and \emph{random-scheduling}. On the other hand, in \emph{grouped
scheduling}, user requests arrive in groups (e.g. as in scientific
calculations or business statistics) and it is required that the scheduler
optimizes the arrangement of tasks according to their resource requirements
and the current state of the infrastructure. This approach outperforms
two competitors: \emph{genetic-algorithm }and \emph{modified-genetic-algorithm}~\citet{Kaur2012}
in terms of makespan. This work is limited in terms of achieved algorithm
scalability since both the dimensions of the state space and the number
of actions depend on the number of used VMs, which could make the
problem difficult to solve in the context of tens of VMs. In fact,
the authors perform the experiments with a maximum of 10 VMs and only
considered homogeneous VMs. A heterogeneous infrastructure would make
it possible to use VMs that best fit the resource requirements of
different types of tasks, which helps to achieve a higher execution
efficiency.

Xiao et al.~\citet{Xiao2017} propose a distributed mechanism for
scheduling independent tasks in the context of hybrid Clouds, i.e.
an infrastructure combining public Clouds and third-party Clouds.
The authors aim to maximize the capacity of the available processing
entities (PE) (physical or virtual machines) by considering a cooperation
scheme between the schedulers of the different Clouds. The approach
guides the scheduling decisions based on experience, and therefore,
the problem of each scheduler is modeled as an MDP and \emph{Q-learning}
is used to obtain the appropriate scheduling policy. For the definition
of the states, the authors consider the task type and the workload
of all the possible units \textendash i.e., PEs or schedulers at different
layers\textendash{} to which tasks can be assigned. The authors use
an aggregation strategy to reduce the state space based on a predefined
granularity parameter. The actions correspond to the selection of
the unit to which a task will be assigned. In this sense, the proposal
could have algorithm scalability problems since the space of actions
depends on the number of units, which might be large in a real Cloud.
Then, the reward is defined as an inversely proportional function
to the response time associated with the scheduling, i.e. the reward
corresponds to the objective of the optimization problem. The results
show that the proposal improves the response time compared to five
state-of-the-art scheduling algorithms~\citet{Xhafa2009}: \emph{opportunistic-load-balancing},
\emph{minimum-execution-time}, \emph{minimum-completion-time}, \emph{switching-algorithm},
and \emph{k-percent-best.}

Duggan et al.~\citet{Duggan2017} propose an RL-based strategy to
schedule migrations of VMs, taking into account the current use of
network resources in a Cloud. The idea is to learn to determine the
most appropriate time to migrate a group of VMs from an overloaded
physical machine to an underloaded one. The proposal aims to reduce
the saturation of network resources during rush hours, as well as
to reduce the migration time of the involved VMs. The authors define
a state space based on (\emph{i}) the bandwidth level, determined
by current usage concerning a threshold (ranging from 0 to 100 in
percentages) and (\emph{ii}) the current direction of the network
traffic considering the previous two time steps of bandwidth utilization
(i.e increasing, decreasing, stable). The actions to be carried out
are to perform or delay the migration of a scheduled group of VMs.
Also, it is used as negative reward based on the total migration delay.
The RL-based algorithm used is \emph{Q-learning}. For comparison purposes,
the authors used an algorithm called \emph{minimum-migration-time}~\citet{Beloglazov2012},
which migrates VMs based on the amount of RAM used. Besides, the authors
define a cost function to evaluate the migrations based on the network
saturation level at the time of migration, the migration duration,
and a penalty in case of waiting. The proposal~\citet{Duggan2017},
in comparison with the minimum-migration-time algorithm, was able
to reduce migration cost, as well as the use of network resources
measured as the extra amount of Gb consumed from the link. The results
also show that the RL-based strategy learns to perform the migrations
when there is less network traffic. Moreover, this strategy contributes
to reducing network saturation during rush hours and reduces the duration
of the migrations.

Soualhia et al.~\citet{Soualhia2018} propose ATLAS+, a MapReduce\emph{-}based\emph{~}\citet{Lee2012}
task scheduler for Hadoop~\citet{Glushkova2019}. ATLAS+ is dynamic,
adaptable and its goal is to minimize task failures, defined as unforeseen
events in the Cloud environment such as data loss in storage systems,
hard-drive failures, and so on. The proposed framework is based on
3 components: (i) a machine learning algorithm (Random Forest) to
predict the probability of task failures, (ii) a dynamic predictor
of possible infrastructure failures and, (iii) a scheduler based on
policies generated by an MDP. For scheduling purposes, the stages
in the life cycle of tasks is modeled (submitted, scheduled, waiting,
running, completed, failed). Then, the possible actions to change
the status of a task are: process, reschedule, or kill the task. The
objective of this proposal is to reduce the task failures to have
a minimum impact on their execution time. To learn the policy a variant
of RL that starts with the \emph{SARSA} algorithm (for further exploration
of the policies) in the first 30 minutes is proposed, and then, \emph{Q-learning}
is used for further exploitation of acquired knowledge. Experiments
show that this proposal outperforms other Hadoop schedulers as \emph{FIFO}
and \emph{Fairy} \emph{Capacity}. Reductions of 59\%, 40\%, and 47\%
in the number of failed tasks, the total execution time, and the task
execution time, respectively, were observed. The approach also reduces
the use of CPU and memory by 22\% and 20\%, respectively.

Dutreilh and Kirgizov~\citet{Dutreilh2011} present VirtRL, an autonomous
solution to the problem of dynamic adaptation of the number of resources
allocated to Cloud applications. VirtRL is based on the \emph{Q-learning}
algorithm. In VirtRL, the number of user requests per second, the
number of VMs assigned to the application, and the average response
time of requests are considered for the definition of the states.
Then, the actions represent the number of VMs to acquire or release
(experimental setup comprises actions bounded between -1 and 10) while
the reward considers the cost of acquiring or maintaining the VMs
and a penalty for Service-Level Agreement (SLA) violations. An SLA
is a commitment between a service provider and a client. Particular
aspects of the service such as quality, availability, and responsibilities
are agreed between the service provider and the service user. Concretely,
the main contribution of this work~\citet{Dutreilh2011} is to include
and integrate into an automatic Cloud autoscaler the following elements,
(i) initialization of the function $Q(s,a)$, (ii) acceleration in
the convergence at regular intervals of observations and, (iii) a
mechanism for detecting changes in the performance model.

Moreover, Ghobaei-Arani et al.~\citet{Ghobaei-Arani2018} propose
an RL-based resource provisioning approach for Cloud service applications.
The \emph{Q-learning} algorithm is used for a decision-making agent
to learn when to add or remove VMs to find a satisfactory compensation
between SLA and costs. The authors define 3 possible states based
on the CPU utilization degree of the infrastructure, i.e., when the
CPU utilization is less than the lower threshold, the state is labeled
as under-utilization. Moreover, when the CPU utilization is greater
than the upper threshold, the state is labeled as over-utilization.
Otherwise, the state is normal-utilization. Then, based on the current
state one out of 3 possible actions is selected: scale in, scale-out
or no-action, which means increasing, reducing, or maintaining the
infrastructure. The reward function R assigns a fixed value for each
state-action pair, while prioritizing Scale-out if the state is under-utilization,
Scale-in if the state is over-utilization, and No-action if the state
is normal-utilization. The performance was evaluated with real workloads
and the approach was compared with 3 state-of-the-art strategies:
Cost-aware-LRM~\citet{Yang2014} and Cost-aware-ARMA~\citet{Roy2011},
which are both based on workload predictions, and DRPM~\citet{Al-Ayyoub2015},
a multi-agent system to monitor and provision Cloud resources. The
results showed that this approach was able to reduce by 50\% the total
cost and increase the use of resources by 12\%.

Dezhabad and Sharifian~\citet{Dezhabad2018} address the automatic
autoscaling of virtualized firewalls in a Cloud. The authors propose
GARLAS, a solution that combines RL with a genetic algorithm and queue
theory. The idea of this work is to determine the number of firewalls
that must be active at all times according to the intensity of the
input load and the proportion of requests that each one handles. This
approach aims to optimize the balance between the firewalls use degree
and compliance with SLA related to system performance (for example,
response time). On the one hand, an automatic autoscaler based on
RL that decides when it is convenient to increase or reduce the number
of active firewalls by dynamically adjusting the system to avoid overloading
or wasting resources, is proposed. On the other hand, a genetic algorithm
is responsible for deciding the appropriate proportion of requests
that each of the firewalls must handle, thus balancing the load to
minimize the system response time. For the RL-based autoscaling problem,
states that consider the current request rate and the number of active
firewalls are defined. The actions consist of increasing, reducing,
or maintaining the number of active firewalls, and the reward is responsible
for penalizing overload or low load states, as well as SLA violations.
Then, through the Q-learning algorithm, it is possible to converge
to an appropriate scaling policy. The proposal was compared with static
strategies (number of fixed firewalls) and rule-based strategies (number
of firewalls varies according to load levels). The results show that
GARLAS was able to significantly reduce the response time of the system
(by more than 80\%) and also offers improvements in the use of resources
(more than 9\%). This is due to a better load balance and a more precise
automatic scaling algorithm.

Horovitz and Arian~\citet{Horovitz2018} present a Q-learning solution
for horizontal scaling that adds initialization and smoothing rules
combined with a utilization rate of states change. In this approach,
a state space reduction method is used by exploiting the monotonic
behavior of the actions taken as a function of the state space (e.g.
utilization). Then, an action space reduction method is used for those
actions that are continuous for a given state. Lastly, an innovative
approach to Q-Learning based auto-scaling is applied, the Q-Threshold
algorithm. The authors define the state space as the current number
of resources allocated to the application. The actions consist of
selecting a utilization threshold from $N$ possible values. The threshold
value serves as bound for a given resource (e.g. CPU utilization,
load, response time, etc). The authors maintain two Q-Tables, one
for the upper threshold and another for the lower threshold. Regarding
the reward, it balances the response time and the average resource
utilization, i.e., the reward tries to meet the SLA while making the
resource utilization as high as possible.Therefore, instead of the
traditional action space where machine addition and removal actions
are used, the thresholds values drive the actions, and the traditional
thresholds are dynamically controlled by the algorithm. Q-Threshold
learns the best resource utilization thresholds in the horizontal
autoscaling problem to derive the optimal policy while abiding by
the SLA and maximizing resource utilization.

Finally, Wei et al.~\citet{Wei2019} propose an approach based on
the Q-learning adjustment algorithm (QAA) to help SaaS providers make
optimal resource allocation decisions in a dynamic and stochastic
Cloud environment. The goal of this work is to reduce renting expenses
as much as possible while providing sufficient processing capacity
to meet customer demands. For this, the authors have considered different
VM pricing models, including on-demand and reserved instances. The
authors consider the following features for the states: the average
customer workload, the number of VMs of each type, and a reference
to the specific time within the workload period. Each action comprises
the number of VMs of each type that will be acquired for the next
execution period. Also, the reward function is calculated based on
the profit that SaaS provider earned by providing service to his end-users,
and on the performance (the gain of application performance), which
depends on the resource utilization levels. If SaaS provider owns
sufficient VM instances to execute customer workloads, a positive
reward will be received. In contrast, a penalty will be obtained if
application processing capacity is lower than the customer's demands.
The value of reward or penalty is related to the distance between
the offered processing capacity and the real customer workload. SaaS
provider keeps learning from previous renting experiences and enriching
its knowledge. This accumulated information can help the provider
know the best choices in different situations and then generate an
efficient renting policy for each decision period. Through a series
of experiments and simulations, the authors evaluate QAA under different
pricing models (on-demand and reserved instances) and compare it with
two other resource allocation strategies: empirically-based adjustment
algorithm (EAA) and threshold-based adjustment algorithm (TAA). EAA
adopts a simple strategy to generate a new renting policy. Since SaaS
provider does not know the upcoming customer workload when making
decisions, the provider adjusts the number of rental VM instances
according to the last workload. TAA is similar to EAA but does not
change the renting policy each time. Only when the difference between
customer workload and processing capacity offered by the SaaS provider
exceeds a specific threshold, a new renting policy will be generated.

\subsubsection{Pure Proposals with Parallel Learning\label{subsec:parallelProposals}}

In this section we describe those works from the literature that use
Model-free techniques in their original formulation (as opposed to
proposals based on DRL or FRL), but with a \emph{parallel} learning
process. The main advantage of parallel learning is that training
time is reduced due to multiple agents simultaneously sharing the
acquired knowledge, i.e. the agents periodically exchange information
thus accelerating convergence towards the optimal policy. In this
way, the table $Q(s,a)$ is updated at a faster pace, so it is possible
to obtain a higher quality policy faster, at the expense of higher
approach design complexity.

Barret et al.~\citet{Barrett2012} propose CloudRL, a method based
on MDP and \emph{Q-learning} for dynamic scaling in IaaS infrastructures,
in response to changes in workload and infrastructure performance.
The states are defined based on the number of user requests, the number
of VMs of each type and region, and the \emph{Coordinated Universal
Time} (CUT). Actions are either requesting, maintaining, or removing
instances, while the reward includes the cost and a penalty in case
of SLA violations. Particularly, the authors introduce a strategy
with multiple agents learning in parallel to mitigate the problem
of long convergence time of \emph{Q-learning. }The long convergence
time of \emph{Q-learning} is due to it does not have a good initial
approximation of $\pi$ and a good initialization $Q(s,a)$. The authors
also suggest that parallel learning is scalable in terms of resource
growth because the number of learning agents can be determined based
on the number of available computational resources. In this sense,
the proposal should also be scalable in terms of the number of user
requests. If we take into account the algorithm scalability in both
dimensions, the state space would grow considerably. As a consequence,
this would have an impact not only on the parallel computing capacity
of the agent but also on the storage capacity to keep accessible information
shared between them and the mechanisms for sharing such information.

Benifa and Dejey~\citet{BibalBenifa2018} present RLPAS, an RL-based
approach for automatically scaling virtualized resources in a Cloud.
The objective of the proposal is to dynamically configure resources
to minimize response time while maximizing resource utilization and
performance. The states are defined based on the number of user requests,
the infrastructure utilization degree (the relationship between acquired
and used VMs), as well as the response time and performance observed
for each task during a pre-determined period. Then, the \emph{scale-up}
(or \emph{scale-down}) actions comprise the number of VMs of each
type that will be acquired (or released) in the next execution period
and \emph{no-action} specify that changes are no required. Moreover,
the reward is based on the relationship between performance (related
to response time, throughput, and SLA violations) and the VMs utilization
degree. This approach, based on the \emph{SARSA} algorithm, reduces
convergence time, and combines parallel learning with an approximation
of the function $Q(s,a)$. This approximation is performed by the
gradient descent method. For experimentation, reference applications
with dynamic workloads were used and RLPAS was compared with the pure
variants of \emph{Q-learning} and \emph{SARSA}, as well as with the
approach proposed in~\citet{Barrett2012} and discussed in the previous
paragraph. RLPAS outperformed its competitors in terms of CPU utilization,
response time, performance (number of requests processed per second),
and convergence time.

Nouri et al.~\citet{Nouri2019} present a decentralized RL-based
technique for responding to volatile and complex arrival of tasks
through a set of simple states and actions. The technique is implemented
within a distributed architecture that cannot only scale up quickly
to meet rising demand but also scale down by shutting down excess
servers to save costs. The states consist of two types of attributes:
system state and application state. System state reflects the level
of utilization of resources of a server such as CPU, and the application
state represents the performance of each application hosted on the
server in terms of metrics such as its response time. To make the
state\textendash space discrete, the system states and application
states are classified into three categories: normal, warning, and
critical. On the other hand, the actions are categorized into two
groups: scale-down and scale-up. Scale-up actions would be suitable
when the system is not able to meet the SLA and needs more computing
resources. In contrast, scale-down actions suit situations in which
the system is in normal condition, and idle resources can be released
to minimize cost. The application actions involve either duplicating,
or creating extra instances of an application, or move, wherein an
application is shifted to a different server with more available resources.
The reward for reaching a state is determined by the summation of
all VM utilities. The utility is based on performance, in terms of
the response time of requests, and cost. In this approach, it is feasible
to share the states, take actions, and receive rewards among the servers
to speed up the learning process. Hence, if a server reaches a state
which has not been observed by itself, it tries to find the knowledge
of the state from the shared knowledge base. In case that the look-up
procedure produces no result, it will take the best possible action
using the learning policy. Furthermore, this procedure allows new
servers to initialize their knowledge database using the existing
shared knowledge. The authors evaluate the decentralized control technique
using workloads from real-world use cases and demonstrate that it
reduces SLA violations while minimizing the cost of infrastructure
provisioning.

\subsubsection{Proposals combined with Neural Networks\label{subsec:combinedNN}}

In this section we describe the works that propose solutions exploiting
neural networks (see sub-appendix~A.3), combining techniques of RL
with Deep Neural Networks (DNN) to mitigate the problem of the state
dimensionality associated with RL-based techniques in its purest variant.
This combination is called Deep Reinforcement Learning (DRL). When
including the use of DNN in the decision-making process, it is important
to consider that the solution becomes more complex, in addition to
the problems inherent to DNN~\citet{Gary2018}. On the one hand,
deep learning models usually include a high number of hyper-parameters
(for example, \emph{learning rate}, \emph{batch size}, \emph{momentum},
and \emph{weight decay})~\citet{Smith2018} and finding the best
configuration for these parameters in a large dimensional space is
not trivial. On the other hand, DNNs require a large volume of data
and consequently a lot of training time. Nevertheless, DRL has proven
to be a promissory technique since it allows working with very complex
state spaces and actions. In this sense, the following works show
a first approach to applying DRL to the area of \LyXZeroWidthSpace \LyXZeroWidthSpace autoscaling
in Clouds.

Liu et al.~\citet{Liu2017} propose a hierarchical framework to solve
two important problems in the context of Cloud Computing: task scheduling
and the management of energy consumption of the infrastructure. The
framework consists of a global decision layer for the scheduling problem
and a local decision layer for distributed energy management in local
PM. At the global framework level, a DRL-based strategy capable of
handling the complex state space and actions that characterize the
problem is proposed. In the definition of the states, infrastructure
information (utilization degree of each PM) and task information (resource
requirements and estimated duration) are represented. The actions
correspond to the assignment of the tasks to some of the existing
PM. Then, the reward is composed of three terms with values negatively
weighted of instantaneous total power consumption, the number of VMs
in the system, and reliability objective function value. The DRL-based
strategy consists of an \emph{offline }construction stage of the neural
network, which represents the correlation between the estimated values
\LyXZeroWidthSpace \LyXZeroWidthSpace of $Q$ and the proposed state-action
pairs. Then, the strategy continues to operate in an \emph{online}
stage of decision-making and learning-based both on \emph{Q-learning}
and the update of the neural network previously trained. On the other
hand, the local level of the framework is responsible for energy management,
with a distributed mechanism to selectively turn on and off the PMs.
This level includes a workload predictor based on a Long Short-Term
Memory (LSTM)~\footnote{This kind of neural networks is widely used for predicting time series
sequences.} neural network~\citet{Hochreiter1997}. Then, an RL-based adaptive
energy manager controls the status of any PM based on the workload
predictions made by the LSTM. The authors use Google server logs for
experiments and the Round-Robin scheduling method for comparisons.
The results show that the RL-based proposal achieves significant energy
savings, as well as the best relationship between latency (defined
as the time between the arrival of a task and its completion) and
energy consumption.

Cheng and Nazarian~\citet{Cheng2018} present DRL-Cloud, an approach
based on DRL for workflow scheduling in Clouds. The objective of this
strategy is to minimize the energy cost from the perspective of public
Cloud providers. For them, the authors propose a two-stage (i.e. resource
provisioning and task scheduling) process based on DRL, which is highly
scalable and adaptable. The definition of the states includes infrastructure
information (CPU and RAM availability) and workflow task information
(deadlines, and CPU/RAM requirements). In the first stage, the actions
correspond with the selection of one of the servers farm for allocating
the task and the possible start time. In the second stage, the actions
represent the selection of a specific server to run the task. The
reward on each stage is computed based on the energy cost increase
relative to the farm or the server, respectively. DLR-Cloud is fully
parallelizable and uses training techniques~\citet{Mnih2015} (\emph{target-network}\footnote{\emph{Target Network}: strategy that proposes the use of a second
\flqq{}objective\frqq{} network, during the training of a DQN, to
calculate the updated values \LyXZeroWidthSpace \LyXZeroWidthSpace of
$Q$. In this way, a more stable training is achieved since the weights
of this second network are updated less frequently than those of the
original network.}\emph{, experience-replay}\footnote{\emph{Experience Replay: }a strategy that proposes to store the agent's
experiences and then use random data for the training of the DQN.
In this way, correlations in the observation sequences are eliminated
and changes in data distribution are smoothed out.}) to accelerate convergence. The proposal was compared with two methods:
Fast and Energy-Aware Resource Provisioning and Task Scheduling (FERPTS)~\citet{Li2017}
and Round-Robin. Results show significant improvements in the reduction
of energy costs, the number of rejected applications (due to deadline
violations), and the execution time.

Wang et al.~\citet{Wang2017} explore the use of RL techniques for
horizontal scaling in the Cloud. The idea of this work is to learn
policies capable of adjusting the infrastructure, achieving a balance
between performance and costs. The states comprise the number of VMs,
two instance-level CloudWatch\footnote{Amazon CloudWatch: \url{https://aws.amazon.com/cloudwatch/}}
metrics (CPU Utilization, Network Packets In), and two elastic load
balancer-level metrics (Latency, Number of Requests). The actions
include increasing and reducing the infrastructure in one or two VMs,
as well as maintaining the current resources. A negative reward is
computed considering the cost of the provisioned resources and a penalization
relative to the CPU utilization (depending on the SLA). The authors
show a preliminary study of 3 strategies of RL: \emph{tabular-Q-learning
}(QL), \emph{deep-Q-network} (DQN) y \emph{double-dueling-Q-network}
(D3QN), first in the CloudSim simulator~\citet{humane2015simulation}
and then in the Amazon Cloud. QL corresponds to a classic variant
of \emph{Q-learning} where the function $Q$ is represented in tabular
form. DQN uses a deep neural network to estimate the $Q$ function.
D3QN is another variant that uses a second neural network (Double
Q-Network~\citet{Wang2015}) to stabilize the training process and
to update the value of the states in a more robust and decoupled form
from the specific actions (\emph{Dueling Q-network}~\citet{Wang2015}).
The study compares a dynamic scaling method based on predefined thresholds
of CPU usage (threshold-based method) with the three above mentioned
methods based on RL. The results show the superiority of the proposed
DRL-based methods. Besides, D3QN significantly outperformed DQN in
terms of the accumulated reward and learning speed.

Tong et al.~\citet{Tong2020} propose a deep Q-learning task scheduling
(DQTS) that combines the advantages of the Q-learning algorithm and
a deep neural network. This approach is aimed at solving the problem
of scheduling workflow tasks in a Cloud and uses the popular deep
Q-learning (DQL) method~\citet{Alam2019}. The goal is to learn policies
capable of adjusting resource utilization while minimizing the makespan.
The approach comprises three layers: task submission layer, deep Q-learning
algorithm layer, and workflow management system layer. The states
comprise a normalized and weighted combination of the current task
processing time and the current accumulated processing time on each
VM. The actions represent the selection of one VM for scheduling the
task and the reward is computed as a positive or negative value considering
the condition of the selected VM (i.e. idle or busy).

Du et al.~\citet{Du2019} propose a DRL approach to make Cloud resource
pricing and allocation decisions for reducing cost. For this, the
authors consider both the VM pricing and placement of VMs in the DRL
model, and time-variant user dynamics, rather than simply assuming
user arrivals are independent and identically distributed. The approach
combines long short-term memory (LSTM) networks with the Deep Deterministic
Policy Gradient (DDPG) method to deal with online user arrivals, and
use a new update method to allow them to work together and to learn
optimal decisions directly from input states. In this approach, for
each VM request, the provider decides the server with available capacity
to allocate the VM and the cost for running the VM on the selected
server. The states are composed of two parts. The first part includes
the current resource availability on all servers and information about
the new VM requests, and the second part includes the states history
encoded by the LSTM. Then, the action space of the DRL agent includes
both discrete actions for server selection and continuous actions
for pricing. The reward, when a user accepts the posted price, is
the payment of the user minus the increased cost of the server due
to running the user VM; otherwise, the reward is 0. The output of
the neural network is divided into two parts: one gives the probability
distribution for choosing among different servers and the other produces
the corresponding unit-time-usage cost on the servers if the requested
VM is to be allocated there. Therefore, the goal of this approach
is to learn a policy that minimizes the costs of the Cloud provider.

\subsubsection{Proposals combined with Fuzzy Logic\label{subsec:combinedFL}}

This section discusses the works that combine RL-based techniques
with elements of Fuzzy Logic (FL) as another alternative to the dimensionality
problem that arises in the purest variants of RL.

Arabnejad et al.~\citet{Arabnejad2017} address the problem of horizontal
resource scaling in the Cloud to reduce application costs and ensure
SLA compliance. The states are defined based on the workload, the
response time, and the number of VMs. The actions are either increasing
(one or two VMs), reducing (one or two VMs), and maintaining the infrastructure.
The reward depends on the number of resources acquired and SLA violations.
In this work, the authors propose and compare two strategies based
on RL and an FL system. The modified versions (\emph{Fuzzy} \emph{Q-learning}
and \emph{Fuzzy} \emph{SARSA}) of the classic RL algorithms can learn
and modify fuzzy scaling rules during execution. Both proposals are
implemented and compared on the OpenStack\footnote{OpenStack: Open Source Platform for Cloud Computing (https://www.openstack.org/).}
Cloud platform. The results show that both proposals can handle different
workload patterns, reduce operational costs, and prevent SLA violations,
with acceptable performance in terms of response time and the number
of used VMs. It is important to note that the authors use an environment
limited to a maximum of 5 VMs to evaluate the strategies together
with high workloads.

Veni and Bhanu~\citet{T.Veni2016} present an approach for vertical
scaling of virtualized resources in Clouds. The proposal is based
on neuro-fuzzy reinforcement learning, combining RL with neural networks
and the approximate reasoning of fuzzy logic. This combination aims
to mitigate the limitations of the most basic variants of RL in a
space of large states. The main objective of this work is to dynamically
configure the resources of each VM based on the current workload to
achieve the maximum overall system performance with minimum resource
utilization. The states are defined according to three VM characteristics
(CPU time, CPU number, and memory size). The action set describes
actions such as increase, decrease, or maintain each configurable
resource of a VM. The reward considers the relationship between the
overall performance of the system and the resource utilization degree.
The proposal was compared with a basic variant of RL (Basic-RL) and
with a strategy that combines RL and neural networks (VCONF~\citet{Rao2009}).
The results show that the proposal achieves significant improvements
in system performance and scalability. Recall that vertical scaling
is limited by the underlying hardware. This means that the characteristics
of the VMs can be improved only as far as resources are available
in the PMs that allocate them. For the scaling problem of applications
in the Cloud, it would be interesting that this proposal combines
vertical scaling with some variant of horizontal scaling to better
mitigate hardware limitations.

\subsection{RL Elements Summary}

This section summarizes the RL elements (states, actions, reward,
and time between actions) considered on each work. Table~\ref{tab:ComparativeSummary-RL}
present (\emph{i}) the features considered for the states definition,
(\emph{ii}) the action space, (\emph{iii}) the model of the environment,
(iv) the discount factor, (v) an informal notion of the reward function,
and (\emph{v}i) the time between actions. Please note that the table
does not include the RL element ``model of the environment'' for
all the surveyed works. The entry is only present for the two works
discussing \emph{model-based} methods. In the context of autoscaling
in Cloud, the states usually reflect information about the situation
of the infrastructure (number of VMs, workload, utilization level,
etc.) as well as the application conditions ( performance, requirements,
etc.). Notice that, some works are focused on scaling actions that
are related to the adjustment of the available resources (e.g. increase,
reduce or maintain the number of VMs). Notice also that other works
are focused on scheduling actions that reflect the selection of a
specific resource where the current task will run. Moreover, different
and complex reward (or penalization) functions have been defined to
optimize multiple objectives (e.g. execution time, economic cost,
resource utilization, SLA, etc.). The time between actions can be
fixed (i.e., 5~min or 1~hour) or variable, e.g. when a new user
request (or a new task or task batch) arrives.

\begin{longtable}[l]{>{\raggedright}p{2cm}>{\raggedright}p{12cm}}
\caption{\label{tab:ComparativeSummary-RL}Unified view of the main RL elements
in each surveyed work.}
\tabularnewline
\midrule 
\textbf{\footnotesize{}Reference} & {\footnotesize{}RL elements}\tabularnewline
\midrule
\endfirsthead
\midrule 
\textbf{\footnotesize{}Reference} & {\footnotesize{}RL elements}\tabularnewline
\midrule
\hline
\endhead
\midrule 
{\footnotesize{}Gar\'i et al., 2019~\citet{Gari2019}} & \textbf{\footnotesize{}States:}{\footnotesize{} The proportion of
long tasks, the maximum parallelism degree, the budget limit, and
the probability of out-of-bid error }\linebreak{}
\textbf{\footnotesize{}Actions:}{\footnotesize{} The possible budget
assignment ratio between spot and on-demand instances. Values: $\left\{ 0.0,0.1,...,1.0\right\} $
}\linebreak{}
\textbf{\footnotesize{}Model of the environment: }{\footnotesize{}Is
an estimation obtained through the execution of a set of heuristic-based
autoscaling methods.}\linebreak{}
\textbf{\footnotesize{}Discount factor: $\gamma\in\{0.1,0.5,0.9\}$}\\
\textbf{\footnotesize{}Reward:}{\footnotesize{} The ratio between
the execution progress and the cost in the last period}\linebreak{}
\textbf{\footnotesize{}Time between actions:}{\footnotesize{} 1 hour}\tabularnewline
\midrule 
{\footnotesize{}Barret et al., 2011~\citet{Barrett2011}} & \textbf{\footnotesize{}States:}{\footnotesize{} A time-stamp within
the workload period (0-23 hour), the resource load (light, moderate
or heavy) and the execution result (failed or successful) }\linebreak{}
\textbf{\footnotesize{}Actions:}{\footnotesize{} The possible execution
plans generated by the GA} {\footnotesize{}solvers. Values: $\left\{ 1,...,N\right\} $}\linebreak{}
\textbf{\footnotesize{}Model of the environment: }{\footnotesize{}Is
an estimation obtained through experience using a Bayesian Model Learning
approach. }\linebreak{}
\textbf{\footnotesize{}Reward:}{\footnotesize{} The cost and a penalization
based on violated SLA}\linebreak{}
\textbf{\footnotesize{}Discount factor: }{\footnotesize{}Unspecified}\\
\textbf{\footnotesize{}Time between actions: }{\footnotesize{}1 hour}\tabularnewline
\midrule 
{\footnotesize{}Peng et al., 2015~\citet{Peng2015}} & \textbf{\footnotesize{}States:}{\footnotesize{} The remainder of the
buffer capacity of each VM (5 discrete levels)}\\
\textbf{\footnotesize{}Actions:}{\footnotesize{} The available VMs
for allocating the current task. Values: $\left\{ 1,...,N\right\} $
}\linebreak{}
\textbf{\footnotesize{}Reward:}{\footnotesize{} 1, 0, or -1, considering
the mean waiting time of recent user request and if the VM selected
was the one with the maximum capacity}\linebreak{}
\textbf{\footnotesize{}Discount factor: }{\footnotesize{}Unspecified}\textbf{\footnotesize{}}\\
\textbf{\footnotesize{}Time between actions: }{\footnotesize{}Time
until the new user request}\tabularnewline
\midrule 
{\footnotesize{}Xiao et al., 2017~\citet{Xiao2017}} & \textbf{\footnotesize{}States: }{\footnotesize{}The task type and
the workload of all the possible units (i.e PMs/VMs or schedulers
at different layers) to which the tasks can be assigned} {\footnotesize{}}\linebreak{}
\textbf{\footnotesize{}Actions:}{\footnotesize{} The available units
for allocating the current task. Values: $\left\{ 1,...,N\right\} $}\linebreak{}
\textbf{\footnotesize{}Reward:}{\footnotesize{} An inversely proportional
function to the response time associated with the scheduling}\linebreak{}
\textbf{\footnotesize{}Discount factor: $\gamma=0.9$}\\
\textbf{\footnotesize{}Time between actions: }{\footnotesize{}Time
until new task arrival}\tabularnewline
\midrule 
{\footnotesize{}Duggan et al., 2017~\citet{Duggan2017}} & \textbf{\footnotesize{}States:}{\footnotesize{} The bandwidth level
(ranging from 0 to 100 in percentages) and the direction of the network
traffic (increasing, decreasing, stable) }\linebreak{}
\textbf{\footnotesize{}Actions:}{\footnotesize{} Wait or migrate a
scheduled group of VMs}\linebreak{}
\textbf{\footnotesize{}Reward:}{\footnotesize{} A penalization based
on the VM migrations delays }\linebreak{}
\textbf{\footnotesize{}Discount factor: $\gamma=0.5$}\\
\textbf{\footnotesize{}Time between actions: }{\footnotesize{}5 minutes}\tabularnewline
\midrule 
{\footnotesize{}Soualhia et al, 2018~\citet{Soualhia2018}} & \textbf{\footnotesize{}States: }{\footnotesize{}The task status. Values:$\left\{ Submitted,Scheduled,Waiting,Executed,Finished,Failed\right\} $}\linebreak{}
\textbf{\footnotesize{}Actions:}{\footnotesize{} Process, reschedule,
or kill the task }\linebreak{}
\textbf{\footnotesize{}Reward:}{\footnotesize{} Unspecified}\linebreak{}
\textbf{\footnotesize{}Discount factor: }{\footnotesize{}Unspecified}\textbf{\footnotesize{}}\\
\textbf{\footnotesize{}Time between actions: }{\footnotesize{}Unspecified}\tabularnewline
\midrule 
{\footnotesize{}Dutreilh and Kirgizov, 2011~\citet{Dutreilh2011}} & \textbf{\footnotesize{}States:}{\footnotesize{} The number of user
requests, the number of VMs allocated to the application, and the
average response time}\linebreak{}
\textbf{\footnotesize{}Actions:}{\footnotesize{} The number of VMs
to acquire or release. Values: \{-1,0,+1,...,+10\}}\linebreak{}
\textbf{\footnotesize{}Reward:}{\footnotesize{} The cost and a penalization
based on violated SLA}\linebreak{}
\textbf{\footnotesize{}Discount factor: $\gamma=0.45$}\\
\textbf{\footnotesize{}Time between actions: }{\footnotesize{}Unspecified}\tabularnewline
\midrule 
{\footnotesize{}Ghobaei-Arani et al., 2018~\citet{Ghobaei-Arani2018}} & \textbf{\footnotesize{}States:}{\footnotesize{} The CPU utilization
level. Values:$\left\{ UnderUtilization,NormalUtilization,OverUtilization\right\} $}\linebreak{}
\textbf{\footnotesize{}Actions:}{\footnotesize{} Reduce (-1), maintain,
increase (+1) the number of VMs}\linebreak{}
\textbf{\footnotesize{}Reward:}{\footnotesize{} A set of precomputed
reward values, prioritizing for each state, the action that leads
to a balanced CPU utilization}\linebreak{}
\textbf{\footnotesize{}Discount factor: }{\footnotesize{}Unspecified}\textbf{\footnotesize{}}\\
\textbf{\footnotesize{}Time between actions: }{\footnotesize{}5 minutes}\tabularnewline
\midrule 
{\footnotesize{}Dezhabad and Sharifian,2018~\citet{Dezhabad2018}} & \textbf{\footnotesize{}States:}{\footnotesize{} The current request
rate and the number of active firewalls}\linebreak{}
\textbf{\footnotesize{}Actions:}{\footnotesize{} Reduce (-1), maintain,
increase (+1) the number of active firewalls}\linebreak{}
\textbf{\footnotesize{}Reward: }{\footnotesize{}A penalization based
on violated SLA and disadvantageous firewall loads (over-load/under-load)
}\linebreak{}
\textbf{\footnotesize{}Discount factor: }{\footnotesize{}$\gamma=0.8$}\textbf{\footnotesize{}}\\
\textbf{\footnotesize{}Time between actions: }{\footnotesize{}5 minutes}\tabularnewline
\midrule 
{\footnotesize{}Horovitz and Arian, 2018~\citet{Horovitz2018}} & \textbf{\footnotesize{}States:}{\footnotesize{} The number of VMs
allocated to the application}\linebreak{}
\textbf{\footnotesize{}Actions:}{\footnotesize{} Predefined utilization
thresholds. Values: $\left\{ 1,...,N\right\} $}\linebreak{}
\textbf{\footnotesize{}Reward:}{\footnotesize{} A balance between
the response time and the average resource utilization}\linebreak{}
\textbf{\footnotesize{}Discount factor: }{\footnotesize{}$\gamma=0.5$}\textbf{\footnotesize{}}\\
\textbf{\footnotesize{}Time between actions: }{\footnotesize{}1 minute}\tabularnewline
\midrule 
{\footnotesize{}Wei et al., 2019~\citet{Wei2019}} & \textbf{\footnotesize{}States:}{\footnotesize{} The average customer
workload, the number of VMs of each type and a time-stamp within the
workload period}\linebreak{}
\textbf{\footnotesize{}Actions:}{\footnotesize{} The number of VMs
of each type to acquire }\linebreak{}
\textbf{\footnotesize{}Reward:}{\footnotesize{} The profit earned
by SaaS provider and the gain of application performance }\linebreak{}
\textbf{\footnotesize{}Discount factor: }{\footnotesize{}$\gamma=0.5$}\textbf{\footnotesize{}}\\
\textbf{\footnotesize{}Time between actions: }{\footnotesize{}1 hour}\tabularnewline
\midrule 
{\footnotesize{}Benifa and Dejey, 2018~\citet{BibalBenifa2018}} & \textbf{\footnotesize{}States: }{\footnotesize{}The number of user
requests, the infrastructure utilization degree, and (for each task)
the response time and throughput}\linebreak{}
\textbf{\footnotesize{}Actions: }{\footnotesize{}The number of VMs
of each type to acquire or release}\linebreak{}
\textbf{\footnotesize{}Reward:}{\footnotesize{} The ratio between
Performance (i.e response time, throughput, and SLA violations) and
VMs utilization degree }\linebreak{}
\textbf{\footnotesize{}Discount factor: }{\footnotesize{}$\gamma=0.9$}\textbf{\footnotesize{}}\\
\textbf{\footnotesize{}Time between actions: }{\footnotesize{}1 minute}\tabularnewline
\midrule 
{\footnotesize{}Barret et al, 2012~\citet{Barrett2012}} & \textbf{\footnotesize{}States:}{\footnotesize{} The number of user
requests, number of VMs of each type and region, a time-stamp }\linebreak{}
\textbf{\footnotesize{}Actions: }{\footnotesize{}Reduce (-1), maintain,
increase (+1) the number of VMs}\linebreak{}
\textbf{\footnotesize{}Reward:}{\footnotesize{} The cost and a penalization
based on violated SLA}\linebreak{}
\textbf{\footnotesize{}Discount factor: }{\footnotesize{}$\gamma=0.85$}\textbf{\footnotesize{}}\\
\textbf{\footnotesize{}Time between actions: }{\footnotesize{}Unspecified}\tabularnewline
\midrule 
{\footnotesize{}Nouri et al., 2019~\citet{Nouri2019}} & \textbf{\footnotesize{}States: }{\footnotesize{}The system state (i.e.,
CPU utilization) and the state of each application (i.e., average
response time)}\linebreak{}
\textbf{\footnotesize{}Actions:}{\footnotesize{} Server-related actions
are $\left\{ Start,Terminate,Find\right\} $; Application-related
actions are $\left\{ Duplicate,Move,Merge\right\} $}\linebreak{}
\textbf{\footnotesize{}Reward:}{\footnotesize{} The summation of VMs
utilities (i.e., response time and cost)}\linebreak{}
\textbf{\footnotesize{}Discount factor: }{\footnotesize{}$\gamma=0.98$}\textbf{\footnotesize{}}\\
\textbf{\footnotesize{}Time between actions: }{\footnotesize{}30 minutes}\tabularnewline
\midrule 
{\footnotesize{}Arabnejad et al., 2017~\citet{Arabnejad2017}} & \textbf{\footnotesize{}States:}{\footnotesize{} The workload, the
response time, and the number of VMs }\linebreak{}
\textbf{\footnotesize{}Actions: }{\footnotesize{}The number of VMs
to acquire or release. Values: \{-2,-1,0,+1,+2\}}\linebreak{}
\textbf{\footnotesize{}Reward:}{\footnotesize{} A penalization based
on violated SLA and the number of VMs}\linebreak{}
\textbf{\footnotesize{}Discount factor: }{\footnotesize{}$\gamma=0.8$}\textbf{\footnotesize{}}\\
\textbf{\footnotesize{}Time between actions: }{\footnotesize{}Unspecified}\tabularnewline
\midrule 
{\footnotesize{}Veni and Bhanu, 2016~\citet{T.Veni2016}} & \textbf{\footnotesize{}States: }{\footnotesize{}The configuration
of each VM (i.e. CPU time, number of CPUs, and memory size)}\linebreak{}
\textbf{\footnotesize{}Actions:}{\footnotesize{} Reduce, maintain
or increase for each configurable resources of a VM} {\footnotesize{}}\linebreak{}
\textbf{\footnotesize{}Reward:}{\footnotesize{} The ratio between
Performance (i.e., response time, throughput, and SLA violations)
and VM utilization degree }\linebreak{}
\textbf{\footnotesize{}Discount factor: }{\footnotesize{}$\gamma=0.1$}\textbf{\footnotesize{}}\\
\textbf{\footnotesize{}Time between actions: }{\footnotesize{}1 minute}\tabularnewline
\midrule 
{\footnotesize{}Wang et al., 2017~\citet{Wang2017}} & \textbf{\footnotesize{}States: }{\footnotesize{}The number of VMs,}
{\footnotesize{}two instance-level Amazon CloudWatch metrics (i.e.,
CPU Utilization, NetworkPacketsIn), and two elastic load balancer-level
metrics (Latency, Number of Requests) }\linebreak{}
\textbf{\footnotesize{}Actions: }{\footnotesize{}The number of VMs
to acquire or release. Values: \{-2,-1,0,+1,+2\}}\linebreak{}
\textbf{\footnotesize{}Reward: }{\footnotesize{}A penalization based
on cost and CPU utilization }\linebreak{}
\textbf{\footnotesize{}Discount factor: }{\footnotesize{}$\gamma=0.99$}\textbf{\footnotesize{}}\\
\textbf{\footnotesize{}Time between actions: }{\footnotesize{}5 minutes}\tabularnewline
\midrule 
{\footnotesize{}Liu et al., 2017~\citet{Liu2017}} & \textbf{\footnotesize{}States: }{\footnotesize{}The infrastructure
state (i.e., utilization degree of each PM) and the task state (i.e.,
resource requirements and estimated duration)}\linebreak{}
\textbf{\footnotesize{}Actions:}{\footnotesize{} The available PMs
for allocating the current task. Values: $\left\{ 1,...,N\right\} $}\linebreak{}
\textbf{\footnotesize{}Reward:}{\footnotesize{} Penalization based
on power consumption, number of VMs, and reliability issues}\linebreak{}
\textbf{\footnotesize{}Discount factor: }{\footnotesize{}$\gamma=0.5$}\textbf{\footnotesize{}}\\
\textbf{\footnotesize{}Time between actions: }{\footnotesize{}Time
until new task arrival}\tabularnewline
\midrule 
{\footnotesize{}Cheng et al., 2018~\citet{Cheng2018}} & \textbf{\footnotesize{}States: }{\footnotesize{}The server state (CPU/RAM
availability) and the task state (deadline and CPU/RAM requirements)
}\linebreak{}
\textbf{\footnotesize{}Actions:}{\footnotesize{} Stage~1: The available
server farms for allocating the current task and the possible start
time. Values: $\left\{ F_{1,1},...,F_{N,T}\right\} $ }\linebreak{}
{\footnotesize{}Stage~2: The available servers within a specific
farm for allocating the current task. Values:$\left\{ S_{1},...,S_{M}\right\} $}\linebreak{}
\textbf{\footnotesize{}Reward:}{\footnotesize{} Stage~1: energy cost
increase in the Farm, Stage~2: energy cost increase in the server}\linebreak{}
\textbf{\footnotesize{}Discount factor: }{\footnotesize{}$\gamma=0.9$}\textbf{\footnotesize{}}\\
\textbf{\footnotesize{}Time between actions: }{\footnotesize{}Unspecified}\tabularnewline
\midrule 
{\footnotesize{}Tong et al., 2020~\citet{Tong2020}} & \textbf{\footnotesize{}States: }{\footnotesize{}Normalized and linear
combination of current task-processing time and current accumulated
processing time for each VM}\linebreak{}
\textbf{\footnotesize{}Actions:}{\footnotesize{} The available VMs
for allocating the current task. Values: $\left\{ 1,...,N\right\} $}\linebreak{}
\textbf{\footnotesize{}Reward:}{\footnotesize{} 1 if the selected
VM is idle and -1 if busy}\linebreak{}
\textbf{\footnotesize{}Discount factor: }{\footnotesize{}$\gamma=0.5$}\textbf{\footnotesize{}}\\
\textbf{\footnotesize{}Time between actions: }{\footnotesize{}Unspecified}\tabularnewline
\midrule 
{\footnotesize{}Du et al., 2019~\citet{Du2019}} & \textbf{\footnotesize{}States: }{\footnotesize{}The current resource
availability of each server, information of the new VM request, and
states history encoded by the LSTM network}\linebreak{}
\textbf{\footnotesize{}Actions: }{\footnotesize{}The probabilities
and prices of selecting the respective servers}\linebreak{}
\textbf{\footnotesize{}Reward:}{\footnotesize{} The profit (i.e. user
payment - cost) if the user accepts the posted price or zero in other
case }\linebreak{}
\textbf{\footnotesize{}Discount factor: }{\footnotesize{}$\gamma=0.99$}\textbf{\footnotesize{}}\\
\textbf{\footnotesize{}Time between actions: }{\footnotesize{}Time
until the new user request}\tabularnewline
\bottomrule
\end{longtable}

\subsection{Classification of the Reviewed Approaches\label{subsec:AnalysisState-of-the-art}}

For the sake of organizing the surveyed related works, we present
three taxonomies that describe the autoscaling problem in Cloud (see
subsection~\ref{subsec:TaxonomyAutoscaling}), the types of executed
applications (see subsection~\ref{subsec:ApplicationTaxonomy}),
and the optimization objectives that are addressed in each work (subsection~\ref{subsec:TaxonomyObjectives}),
respectively. Then, in subsection~\ref{subsec:ComparativeAnalysis},
an in-depth comparative analysis based on the surveyed works and defined
taxonomies is presented.

\subsubsection{Taxonomy of Autoscaling Problems\label{subsec:TaxonomyAutoscaling}}

\begin{figure}[H]
\begin{centering}
\includegraphics[scale=0.6]{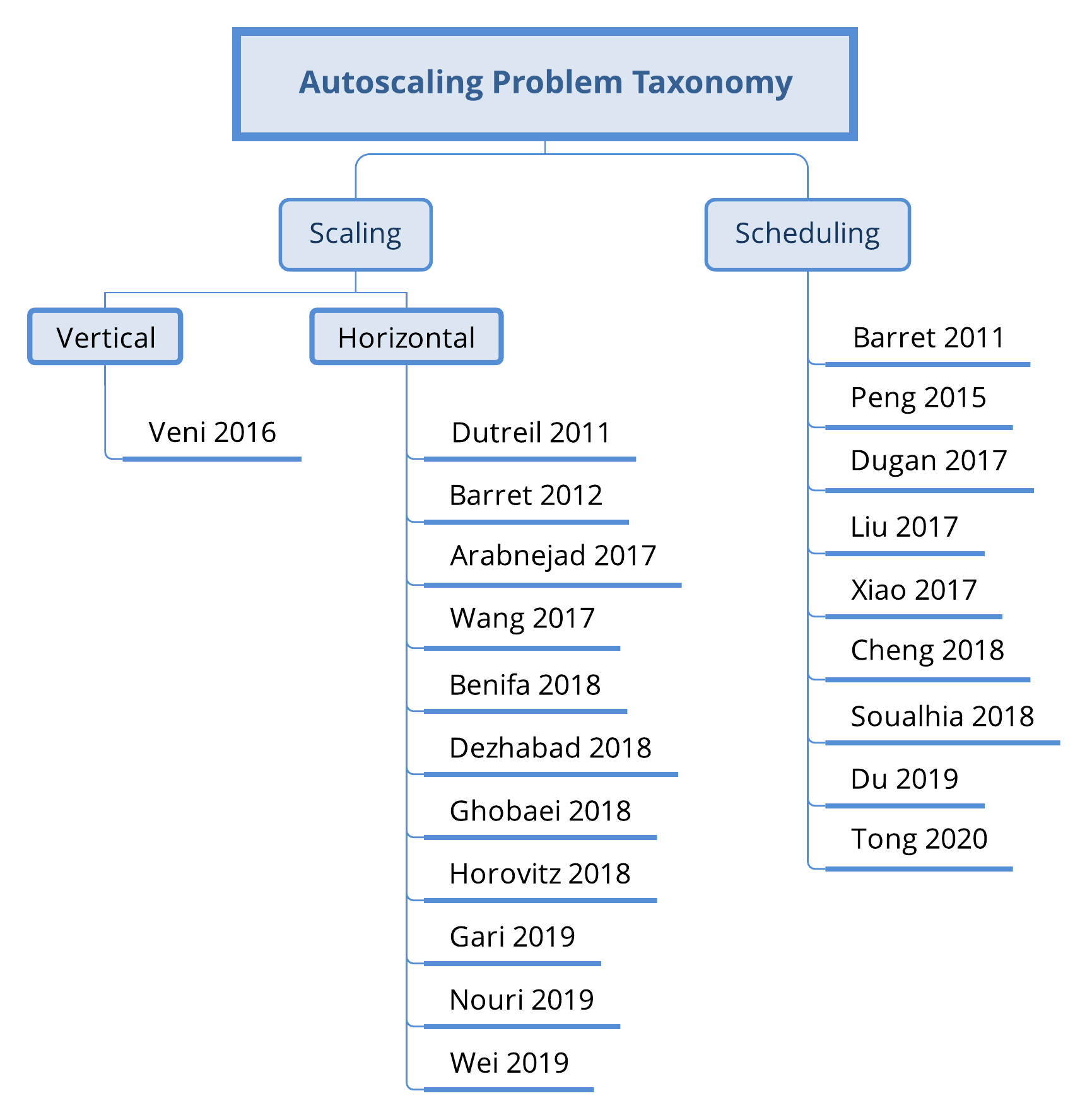}
\par\end{centering}
\caption{\label{fig:AutoscalingTaxo}Taxonomy of autoscaling problems in Cloud
addressed with RL techniques.}
\end{figure}

To identify the specific problem in each surveyed work, in Figure~\ref{fig:AutoscalingTaxo}
we present a taxonomy of the main addressed problems in the context
of autoscaling in Cloud, where the scaling and the scheduling component
are dealt with as two different complex optimization problems with
particular characteristics that we will see next. However, in both
problems, it is intended to make the most out of the available resources
to achieve the best performance regarding the objectives to be optimized.
The optimization objectives (for example, execution time and cost)
guide the search of possible policies, determining when one policy
is more convenient than the other. On the other hand, the constraints
(for example, response time less than 5 seconds) reduce the search
space to include only acceptable policies, leaving out those that
do not allow compliance with the constraints defined for the problem.
Besides, there are also SLAs, which represent an agreement between
the user and the service provider. This agreement defines which aspects
must be respected in the quality of the provided service and also
represent the constraints for the problem. Although optimization objectives
and constraints may have a non-empty intersection, it is important
to consider that they have different roles within the optimization
problem.

The \emph{scaling problem} consists of determining the appropriate
decisions behind provisioning and releasing virtualized resources,
according to the fluctuations in their demands, and to achieve more
efficient use of them. In other words, an automatic mechanism is required
to increase or reduce the infrastructures based on the current resource
needs and taking advantage of the elasticity inherent to Clouds.

Scaling can be classified as horizontal scaling or vertical scaling~\citet{Spinner2014}.
In \emph{horizontal scaling}, the number of VMs assigned to an application
across the infrastructure is increased or reduced. The idea is to
provide the application with the appropriate number of VMs according
to the needs at a given time, to fully exploit the capabilities of
parallelization in high demand periods and reduce the number of resources
in low demand periods. This type of scaling requires spending time
for initialization of the VMs and may not be appropriate for all types
of applications, for example, database-oriented applications where
splitting and distributing data is not trivial. In \emph{vertical
scaling}, the resource settings (CPU, memory, I/O) are dynamically
updated, increasing or reducing the VMs capabilities, without hindering
the execution units where VMs are running. The idea is to constantly
provide the VMs with the necessary capabilities and update them again
when they are no longer required. This type of scaling usually requires
less time for resource configuration (less than 0.5 seconds) than
the time required for horizontal scaling (5 minutes)~\citet{Spinner2014}.

On the other hand, the \emph{scheduling problem }in the Cloud consists
of automatically determining the appropriate decisions about when
and/or where a task should be executed to achieve the greatest possible
efficiency within the limits defined by the existing constraints.
From a temporal perspective, this problem decides which is the most
appropriate time for the execution of a task. In the context of workflow
scheduling, this type of decision is primarily subject to the order
established by the structure of dependencies between tasks. The scheduling
problem can be addressed from the perspective of prioritizing the
execution of critical tasks~\citet{Monge2017} because of the impact
they have on the workflow makespan. On the other hand, from a spatial
perspective, the problem of scheduling aims to make the connection
between each workflow task and the most appropriate resources to perform
their execution. For this, the scheduler usually uses information
related to the characteristics of the tasks and the resources available
to support the decision-making process. When the scheduler does not
have information about the characteristics of the tasks, estimations
are used, usually for task durations.

\subsubsection{Taxonomy of Application Types\label{subsec:ApplicationTaxonomy}}

\begin{figure}[H]
\begin{centering}
\includegraphics[scale=0.6]{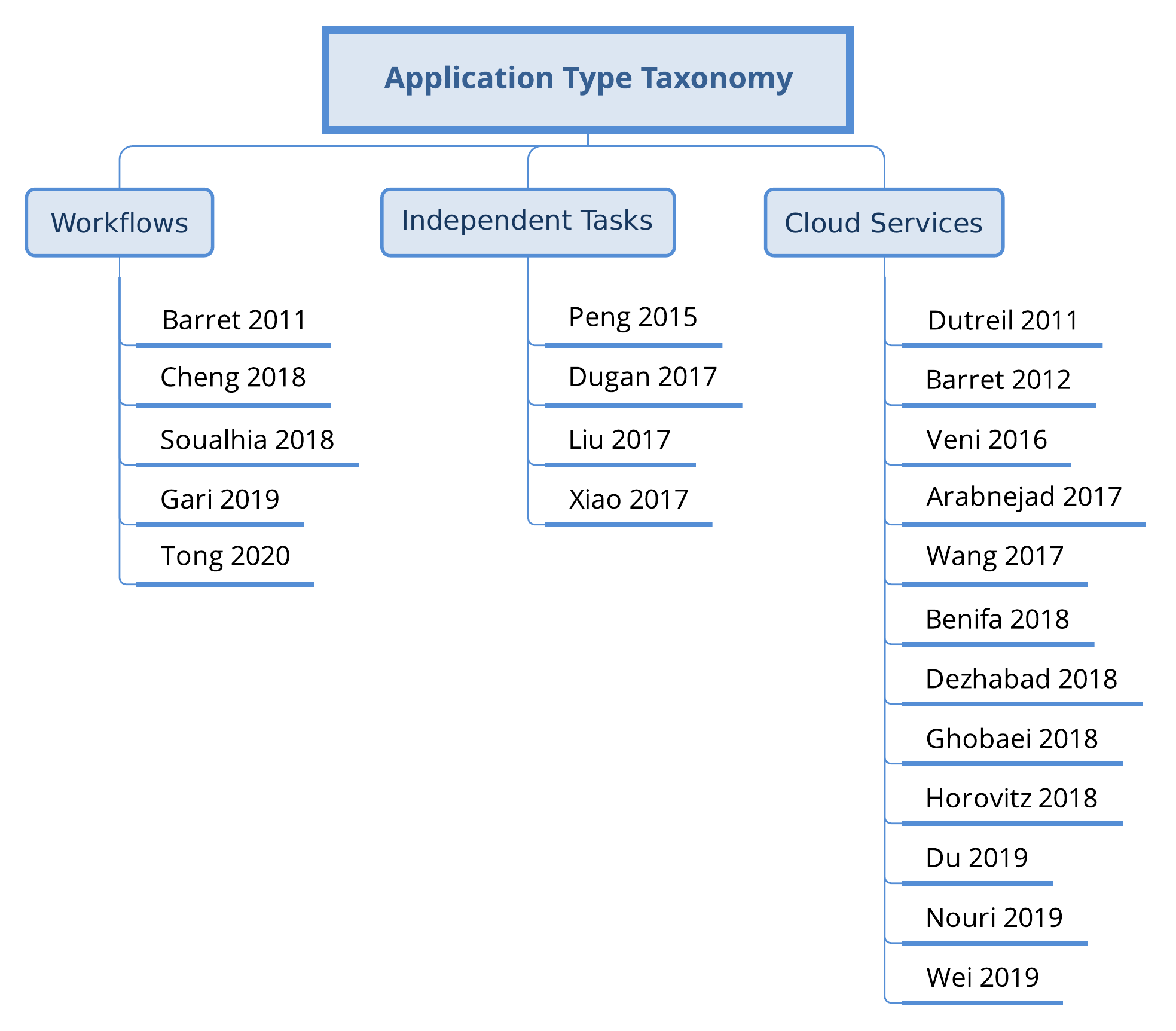}
\par\end{centering}
\caption{\label{fig:ApplicationsTaxo}Taxonomy of Application Types.}
\end{figure}

In the context of Cloud Computing, we can find different types of
applications that need to run efficiently. The reason can be because
the applications involve many hours/days of computation, because they
require many resources, or because immediate results are needed for
multiple user requests. All these distinctive elements are the ones
that also determine the most convenient type of strategy for making
decisions when performing the autoscaling process. In this sense,
it is important to classify the applications and analyze their characteristics.
Figure~\ref{fig:ApplicationsTaxo} shows a taxonomy according to
the application types (see Section~\ref{sec:Applications}) targeted
by the works analyzed in the previous sections.

\emph{Workflow applications }are increasingly used for modeling complex
scientific experiments. Workflows are usually composed of hundreds
or thousands of computing-intensive and/or data-intensive tasks with
different durations and resource requirements. The dependencies between
workflow tasks determine the order in which they must be executed
since a task cannot begin its execution until all the tasks on which
it depends have been completed. The complex structures of dependencies
between workflow tasks determine a variable workload during execution.
This variability is evident when many tasks can be executed in parallel
(high demand for resources). In other cases, bottlenecks may occur,
and it is necessary to wait for the completion of some tasks before
starting the execution of other tasks (low demand for resources).
In this sense, the dependencies between workflow tasks add an important
degree of difficulty to the autoscaling problem. When the workflow
structure is known in advance, it is possible to use workload estimates
to support proper decision-making in the autoscaling process.

Other applications are composed of \emph{independent tasks}. In some
cases, the tasks are randomly initialized by the users so they arrive
individually and continuously, while in other cases the tasks arrive
in batches. When tasks arrive individually and a quick response is
required, the autoscaling decisions are based only on the characteristics
of the current task and on the state of the infrastructure. In this
sense, autoscaling strategies are intended to provide immediate responses.
On the other hand, when the tasks arrive in batches, the number of
options to consider increases exponentially, so strategies that search
in the space of possible solutions are preferred. An example of this
type of applications in the scientific field is the parameter sweep
experiments (PSE), a very popular way of conducting simulation-based
experiments, used by scientists and engineers, through which the same
application code is run several times with different input parameters
resulting in different output data~\citet{Makris2003,GarciaGarino2012b}.
PSEs involve large-scale computer modeling and simulation and often
requires large amounts of computer resources to satisfy the ever-increasing
resource-intensive nature of their experiments. Running PSEs involves
managing many independent tasks. It is important to mention that,
moreover, an independent task application can be considered as a special
type of workflow where there are fictitious start and end tasks, and
all other tasks are intermediate tasks within the workflow structure~\citet{Monge2018}.

The third type of application is \emph{Cloud services}. These applications
represent a software product running on Cloud, that can be accessed
through the Internet either with a Web browser, a mobile application,
or via an API. For example, applications for Big Data Analytics enable
data scientists to tap into any organizational data to analyze it
for patterns and insights, find correlations, make predictions, forecast
future crises, and help in data-backed decision making. Cloud services
make mining massive amounts of data possible by providing higher processing
power and sophisticated tools. Generally, each Cloud service application
is composed of one or more services that together perform the functions
of the application. In this type of applications, the response time
to user's requests is critical. The applications must be scalable
in managing the number of requests, which usually generate high demand
peaks. Although anatomically these applications do not always strictly
comply with a workflow-like structure, we also consider them in this
survey as these are heavily used in practice to execute resource-intensive
models and data analyses on Cloud infrastructures.

Considering the different Cloud service models (SaaS, PaaS, and IaaS)
described in Section~\ref{sec:CloudComputing}, it is interesting
to highlight that from the surveyed works, Cloud service applications
are designed under the SaaS model, while workflows and independent
tasks based works are intended for IaaS and PaaS models.

\subsubsection{Taxonomy of Optimization Objectives\label{subsec:TaxonomyObjectives}}

\begin{figure}[H]
\begin{centering}
\includegraphics[scale=0.6]{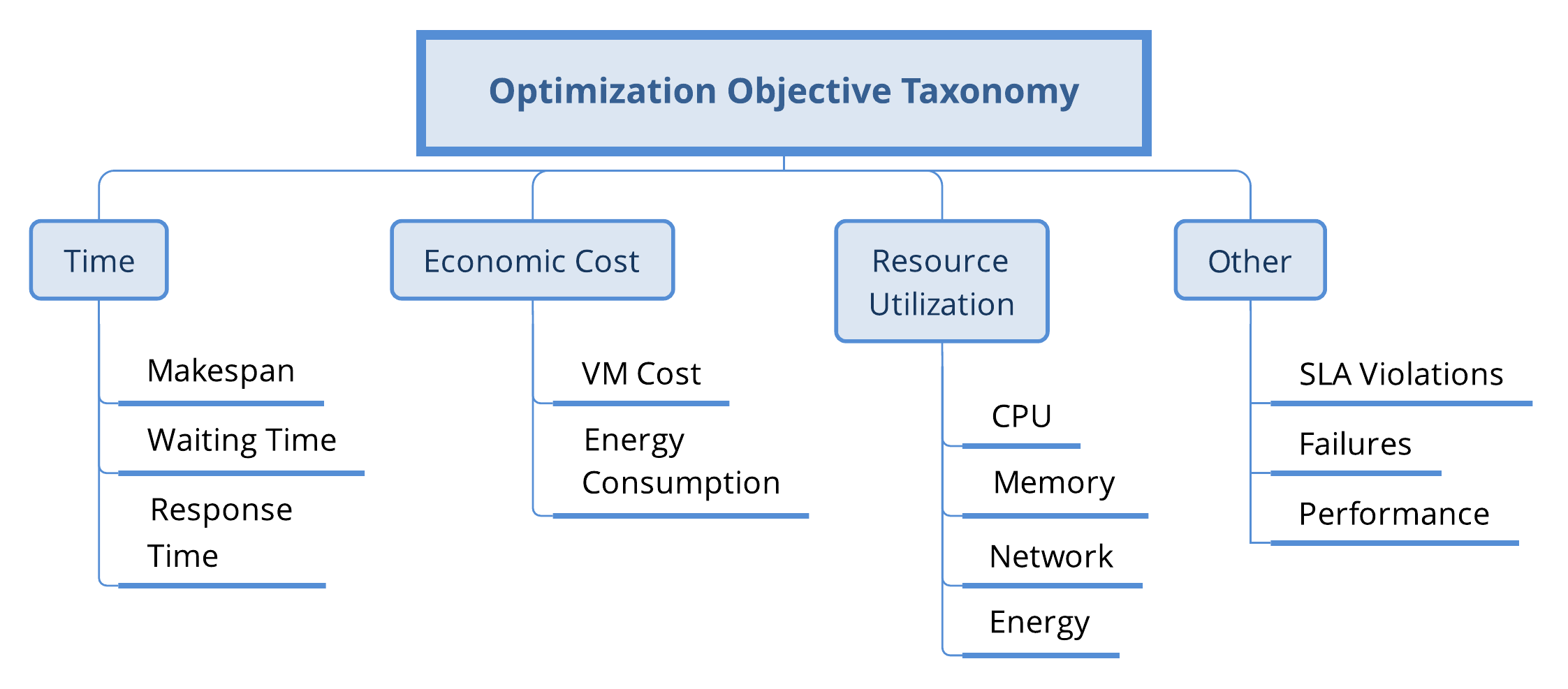}
\par\end{centering}
\caption{\label{fig:TaxoObjetivos}Taxonomy of Optimization Objectives.}
\end{figure}

When optimizing the execution of applications in the Cloud, different
objectives have been addressed. It is very important to identify these
objectives because they represent the direction in which the efforts
will move in guiding the search among possible solutions. The optimization
of multiple objectives is also a recurring issue, and in some cases,
the objectives conflict with each other, as in the classical trade-off
between time and cost in paid Clouds~\citet{Monge2020}. Figure~\ref{fig:TaxoObjetivos}
presents a taxonomy of the objectives targeted in the surveyed works.

First, there are three objectives related to \emph{time}. On the one
hand, it is important to optimize the \emph{makespan} in applications
consisting of compute-intensive or long-duration tasks, as is usually
the case of workflow applications. Note that optimizing workflow makespan
is more complex than optimizing the execution time of each of the
tasks that compose a workflow since the makespan also depends on the
order in which these tasks are executed. On the other hand, the \emph{waiting
time} refers to the time between a task is submitted and actually
begins to execute. The waiting time is usually due to overloaded infrastructures
with high demand peaks and it is of special interest in applications
that process independent tasks. In the case of workflows, the waiting
time is also understood as the delay in starting the execution of
a task after the execution of all the tasks on which this task depends.
For workflow applications, the waiting time is also indirectly optimized
since it impacts the makespan. Finally, in the context of \emph{Cloud
services}, the \emph{response time }that represents the delay in the
response to a user request is usually optimized. This delay may be
due to the overload of the application that needs to be scaled. Both
for workflow applications and independent tasks, the response time
is considered as the sum of the waiting and execution times.

Secondly are the objectives associated with the\emph{ economic cost}.
From the perspective of IaaS users, it is of special interest to reduce
the economic cost associated with the use of VMs (referred to \emph{VM
cost}). This cost is determined by the pay-per-use schemes defined
by public Cloud providers. Each instance type has an assigned price
(usually per hour of use) according to its computational performance.
In this sense, the optimization can be achieved through more efficient
use of the acquired infrastructure, reducing the computation time
required to execute the application. It is also possible to reduce
the execution cost with a dynamic and adequate selection of the VM
types and the most convenient pricing models. For example, the use
of spot or reserved instances represents an interesting opportunity.
Then, from the perspective of public Cloud providers, it is important
to reduce the costs associated with \emph{energy consumption}, due
to their environmental responsibility. The rapid growth of energy
consumption and CO2 emission of Cloud infrastructures has become a
key environmental concern~\citet{Garg2011,Beloglazov2011,Beloglazov2012}.
Therefore, solutions focused on energy efficiency are required to
ensure that the Cloud computing model is sustainable from an environmental
perspective. On the other hand, by reducing expenses on electric bills,
public Cloud providers can also increase their profit margins.

Third, an objective group common to all types of applications are
those that describe the \emph{resource utilization} degree. It is
common to optimize the use of \emph{CPU}, \emph{memory,} and \emph{network}.
Also, in some cases, it is sought to reduce the \emph{number of used
VMs,} and consequently, the computational cost (memory and extra CPU)
associated with their administration, and the \emph{energy consumption.}

Another objective that is evidenced in the literature is to maximize
the system's \emph{performance}. Performance is usually defined considering
the number of user requests served per second, being of special interest
in service applications. For independent tasks and workflows applications,
it is also important to minimize the \emph{failures} of the tasks,
which may be due to hardware or software problems in the infrastructure,
or particularly, the failures associated with the use of unreliable
VM instances (spots or preemptible ones). Finally, the \emph{SLA Violations
}are considered. In many surveyed works, the objective of minimizing
the number of SLA violations agreed between the service provider and
the users are present.

\subsubsection{Comparative Analysis\label{subsec:ComparativeAnalysis}}

Table~\ref{tab:ComparativeSummary} shows a summary of the analyzed
works concerning the taxonomies defined above. First, the works in
terms of the applied RL technique (RL Technique column, see Figure~\ref{fig:TaxoTechniques})
are classified. Furthermore, the RL algorithms used in each case (RL
Algorithm column) is shown. Then, the works based on the specific
addressed problem (Problem column, see Figure~\ref{fig:AutoscalingTaxo}),
the Optimization Objectives (Objectives column, see Figure~\ref{fig:TaxoObjetivos}),
and the Application Type (Application column, see Figure~\ref{fig:ApplicationsTaxo})
are classified, respectively.

\begin{sidewaystable}
\caption{Comparative summary of the relevant works in the literature that apply
RL techniques for Cloud autoscaling.\label{tab:ComparativeSummary}}

\begin{tabular*}{1\columnwidth}{@{\extracolsep{\fill}}llllll}
\toprule 
{\small{}Reference} & {\small{}RL Technique} & {\small{}RL Algorithm} & {\small{}Problem} & {\small{}Objective(s)} & {\small{}Application Type}\tabularnewline
\midrule
\midrule 
{\small{}Gar\'i et al., 2019~\citet{Gari2019}} & {\small{}Model-based} & {\small{}\valueIteration} & {\small{}\scalingH{}} & \emph{\small{}Time, Cost} & {\small{}\wf}\tabularnewline
{\small{}Barret et al., 2011~\citet{Barrett2011}} & {\small{}Model-based} & {\small{}\valueIteration} & {\small{}\sched} & \emph{\small{}Time, Cost} & {\small{}\wf}\tabularnewline
{\small{}Peng et al., 2015~\citet{Peng2015}} & {\small{}Model-free, Pure-Sequential} & {\small{}\qLearning} & {\small{}\sched} & \emph{\small{}Time} & {\small{}\tIndep}\tabularnewline
{\small{}Xiao et al., 2017~\citet{Xiao2017}} & {\small{}Model-free, Pure-Sequential} & {\small{}\qLearning} & {\small{}\sched} & \emph{\small{}Time} & {\small{}\tIndep}\tabularnewline
{\small{}Duggan et al., 2017~\citet{Duggan2017}} & {\small{}Model-free, Pure-Sequential} & {\small{}\qLearning} & {\small{}\sched} & \emph{\small{}Time, Res. Utilization} & {\small{}\tIndep}\tabularnewline
{\small{}Soualhia et al., 2018~\citet{Soualhia2018}} & {\small{}Model-free, Pure-Sequential} & {\small{}\qLearning{} \& \sarsa} & {\small{}\sched} & \emph{\small{}Failures} & {\small{}\wf}\tabularnewline
{\small{}Dutreilh and Kirgizov, 2011~\citet{Dutreilh2011}} & {\small{}Model-free, Pure-Sequential} & {\small{}\qLearning} & {\small{}\scalingH} & \emph{\small{}Cost, SLA} & {\small{}\aCloud}\tabularnewline
{\small{}Ghobaei-Arani et al., 2018~\citet{Ghobaei-Arani2018}} & {\small{}Model-free, Pure-Sequential} & {\small{}\qLearning} & {\small{}\scalingH} & \emph{\small{}Cost, SLA} & {\small{}\aCloud}\tabularnewline
{\small{}Dezhabad and Sharifian, 2018~\citet{Dezhabad2018}} & {\small{}Model-free, Pure-Sequential} & {\small{}\qLearning} & {\small{}\scalingH} & \emph{\small{}Res. Utilization, SLA} & {\small{}\aCloud}\tabularnewline
{\small{}Horovitz and Arian, 2018~\citet{Horovitz2018}} & {\small{}Model-free, Pure-Sequential} & {\small{}\qLearning} & {\small{}\scalingH} & \emph{\small{}Res. Utilization, SLA} & {\small{}\aCloud}\tabularnewline
{\small{}Wei et al., 2019~\citet{Wei2019}} & {\small{}Model-free, Pure-Sequential} & {\small{}\qLearning} & {\small{}\scalingH} & \emph{\small{}Time, Performance} & {\small{}\aCloud}\tabularnewline
{\small{}Benifa and Dejey, 2018~\citet{BibalBenifa2018}} & {\small{}Model-free, Pure-Parallel} & {\small{}\sarsa} & {\small{}\scalingH} & \emph{\small{}Time, Res. Utilization, Performance, SLA} & {\small{}\aCloud}\tabularnewline
{\small{}Barret et al., 2012~\citet{Barrett2012}} & {\small{}Model-free, Pure-Parallel} & {\small{}\qLearning} & {\small{}\scalingH} & \emph{\small{}Resource Utilization, SLA} & {\small{}\aCloud}\tabularnewline
{\small{}Nouri et al., 2019~\citet{Nouri2019}} & {\small{}Model-free, Pure-Parallel} & {\small{}\qLearning} & {\small{}\scalingH} & \emph{\small{}Cost, SLA} & {\small{}\aCloud}\tabularnewline
{\small{}Arabnejad et al., 2017~\citet{Arabnejad2017}} & {\small{}Model-free, FRL} & {\small{}\qLearning{} \& \sarsa} & {\small{}\scalingH} & \emph{\small{}Time, Res. Utilization, SLA} & {\small{}\aCloud}\tabularnewline
{\small{}Veni and Bhanu, 2016~\citet{T.Veni2016}} & {\small{}Model-free, FRL} & {\small{}\sarsa} & {\small{}\scalingV} & \emph{\small{}Time, Res. Utilization, Performance, SLA} & {\small{}\aCloud}\tabularnewline
{\small{}Wang et al., 2017~\citet{Wang2017}} & {\small{}Model-free, DRL} & \emph{\small{}Deep}{\small{} \qLearning} & {\small{}\scalingH} & \emph{\small{}Cost, Resource Utilization} & {\small{}\aCloud}\tabularnewline
{\small{}Liu et al., 2017~\citet{Liu2017}} & {\small{}Model-free, DRL} & \emph{\small{}Deep}{\small{} \qLearning} & {\small{}\sched} & \emph{\small{}Time, Resource Utilization} & {\small{}\tIndep}\tabularnewline
{\small{}Cheng and Nazarian, 2018~\citet{Cheng2018}} & {\small{}Model-free, DRL} & \emph{\small{}Deep}{\small{} \qLearning} & {\small{}\sched} & \emph{\small{}Cost} & {\small{}\wf}\tabularnewline
{\small{}Tong et al., 2020~\citet{Tong2020}} & {\small{}Model-free, DRL} & \emph{\small{}Deep}{\small{} \qLearning} & {\small{}\sched} & \emph{\small{}Time, Resource Utilization} & {\small{}\wf}\tabularnewline
{\small{}Du et al., 2019~\citet{Du2019}} & {\small{}Model-free, DRL} & \emph{\small{}Deep}{\small{} \qLearning} & {\small{}\sched} & \emph{\small{}Cost} & {\small{}\aCloud}\tabularnewline
\bottomrule
\end{tabular*}
\end{sidewaystable}

From the analysis of the characteristics of the surveyed works in
Table~\ref{tab:ComparativeSummary}, the following observations are
highlighting.

Regarding the applied \emph{RL techniques}, most of the works (19/21)
have proposed solutions in the Model-free category and only two works
in the Model-based category~\citet{Gari2019,Barrett2011}, which
is convenient in a context where changes in the environment dynamics
are likely to occur. In this sense, the proposals for \emph{online}
learning seem to be more adequate because they can adapt to these
changes. Besides, there are Cloud autoscaling proposals that attempt
to address classical problems inherent to RL, such as (a) the management
of large state spaces~\citet{T.Veni2016,Arabnejad2017,Liu2017,Wang2017,Cheng2018},
(b) the reduction of training time~\citet{Barrett2012,BibalBenifa2018,Nouri2019},
(c) the poor initial performance~\citet{Dutreilh2011} and (d) the
slow convergence~\citet{Dutreilh2011}.

Regarding the \emph{RL} \emph{algorithms} (particularly in the 19
works in the Model-free category), \emph{Q-learning} (15/19) predominates
over \emph{SARSA} (2/19). Note also that two works are using both
algorithms, i.e., Q-learning and SARSA. However, in most cases, the
selection of \emph{Q-learning} or \emph{SARSA} is not much argued.
It is interesting to note that one of the works~\citet{Soualhia2018}
proposes a combined solution that begins learning with \emph{SARSA}
for further exploration in the policy space and then continues with
\emph{Q-learning} to ensure a more direct convergence towards an appropriate
policy. On the other hand, in~\citet{Arabnejad2017} the authors
combine the use of FL with RL. In this work, modified versions of
both algorithms (Fuzzy-\emph{Q-learning} and \emph{Fuzzy-SARSA}) are
used but no significant performance differences are obtained.

Regarding the \emph{problem}, it can be seen that both scaling (12/21)
and scheduling (9/21) have been the subject of study from the RL perspective.
For the scaling problem, most of the works focus on horizontal scaling,
which means that there is a niche area to explore, i.e. vertical scaling
via RL. No proposals that solve both problems together (scaling and
scheduling) from the perspective of RL have been proposed.

Regarding the \emph{optimization objectives}, in most works, the optimization
of multiple objectives is pursued. Concretely, most of the surveyed
works have proposed to reduce time and costs, to achieve a more balanced
resource usage, to improve performance, and to reduce failures as
well as the violation of restrictions.

Regarding the \emph{application type}, it is observed that service
applications are more associated with the scaling problem. This may
be related to the nature of these types of applications, which need
to be able to scale to the fluctuating demands generated by multiple
user requests while maintaining adequate performance and minimizing
cost. On the other hand, both independent tasks and workflows applications
are more associated with the scheduling problem, assuming a fixed
infrastructure. For this type of applications, usually intensive in
terms of computation power or data processing, it is intended to distribute
the execution of the tasks in the available resources to maximize
efficiency in terms of time, cost, and use of resources. In this sense,
it should be noted that five proposals consider workflows per se or
the above defined special type of workflow composed of independent
tasks (four works), and therefore, workflows are a type of application
that is still to be further explored in the area.

\section{Discussion\label{subsec:Discussion}}

This section analyzes the limitations and scope of the previously
surveyed related works. First, the limitations regarding the type
of addressed problem, the type of application, and the optimization
objectives are analyzed. Then, other more theoretical limitations,
related to the RL techniques used in the proposals, are discussed.
Finally, open problems and ongoing developments in the area are also
discussed.

\subsection{Limitations Related to the Autoscaling Problem Formulation, Application
Type, and Addressed Objectives\label{subsec:Limitations-General}}

\emph{\flqq{}Scaling and scheduling issues are addressed independently\frqq{}}.
There are no works that aim to solve both issues together from the
RL perspective. Especially, for workflows applications and independent
tasks, both problems are interrelated and have an impact on the execution
efficiency, so it is important to design proposals that address both
problems. Since the decisions regarding scaling and scheduling are
different (for scaling, the actions are related to the reconfiguration
of the infrastructure and for scheduling, the actions respond to where
and/or when the tasks will be executed) it could be necessary to define
different models and/or the combination of different techniques to
address both problems.

?\emph{Mix of works with workflow applications}?. Among the 21 analyzed
works, it should be noted that few proposals consider pure workflow
applications (five works) or independent task applications (four works).
Besides, most of these works focus only on scheduling and not on scaling,
except for the work in~\citet{Gari2019}. However, the approach in~\citet{Gari2019}
does not perform scaling purely with RL since it also uses a heuristic-based
autoscaling strategy. Most scaling proposals are still based on service
applications. It is important to note that there is a difference in
the nature of workflow applications (long-term tasks, data-intensive
or compute-intensive tasks, high-parallelism, or bottleneck stages)
compared to service applications (generally short tasks under high
demand peaks). Then, the strategies designed for service applications
are not the most suitable ones for workflows in general, since they
try to optimize processes of a different nature. In this sense, it
is necessary to expand the study on the application of RL techniques
in the context of scaling for the efficient execution of workflows
in Clouds.

?\emph{For the scaling problem, the particular characteristics of
the application about the type of required VM are not considered}?.
Particularly, most of the works define the characteristics of the
environment based mainly on the state of the infrastructure, the number
of available VMs and/or the resource utilization degree, and so on.
Some works include information regarding the state of the application
execution for evaluating the workload. Besides, several authors use
homogeneous infrastructures~\citet{Dutreilh2011,Cheng2018}. Although
using homogeneous infrastructures is a common choice in HPC on the
Cloud, in many cases, using heterogeneous infrastructures leads to
better time and cost optimizations. From the works where the authors
have considered the use of heterogeneous infrastructures~\citet{Ghobaei-Arani2018,BibalBenifa2018,Barrett2012},
only one work~\citet{BibalBenifa2018} represents in the actions
of the model the selection of different types of VMs. The surveyed
works are mostly based on service applications, where the characteristics
of acquired VMs may not be relevant. Conversely, for workflow and
independent-task applications, the characteristics of acquired VMs
are very important since they are usually composed of long duration
tasks that are intensive in computation/data. Therefore, it is necessary
to consider the requirements of the tasks in terms of CPU, memory,
and data transfer, to determine the type of VM where tasks execution
is viable and/or more efficient.

\emph{\flqq{}Different price models are not considered}?\emph{. }Most
of the work in the area focuses on the classical payment scheme, where
the provider defines a fixed price for each VM type charged hourly.
Cloud flexibility is also found in the different options that providers
offer in terms of price models. For example, Amazon spot instances,
whose price fluctuates according to existing demand, have significant
cost reductions (up to 90\% in some cases), compared to the fixed
price model. Considering that the economic cost is one of the main
objectives of optimization in this type of problem and it is also
usually present in the SLA, it is interesting to exploit the Cloud
options in terms of the different price models.

\subsection{Limitations Related to the RL Techniques\label{subsec:Limitation-Techniques}}

\emph{Model-based techniques:} \emph{\flqq{}A perfect model of the
environment is required\frqq{}}. This is one of the main limitations
of the Model-based methods since in many problems the actual distribution
of the transition probabilities between the states is unknown~\citet{Sutton:2018}.
Even in dynamic environments, these probabilities could change over
time. Estimates of the function $P(s,a)$ are usually used, but it
is necessary to take into account that the quality of the obtained
policy depends directly on the quality of these estimates. In the
works surveyed in this category~\citet{Barrett2011,Gari2019}, to
obtain an estimation of $P(s,a)$, the information generated by multiple
previous executions of Cloud applications is used. Although major
public Cloud providers (Amazon, Microsoft, and Google) have access
to a large amount of information regarding executions, and such information
could be used to generate this type of estimated models~\citet{Gari2019},
complexity must be considered for determining the type and amount
of information that should be used to avoid the classic problems of
over-fitting and under-fitting when approximating functions.

\emph{Model-based techniques:} \emph{\flqq{}Offline policies might
no longer be adequate due to changes in the dynamics of the environment\frqq{}}.
The fact that in Model-based approaches the policy is learned \emph{offline}
from a predefined fixed model does not allow it to adjust to changes
in the dynamics of the environment. In the context of Cloud, a change
in instance prices represents a possible cause of variation in the
environment dynamics. A clear example of this is the price fluctuations
of spot instances. When the execution cost is considered as an optimization
objective, the previously computed policy could no longer be adequate
since the learning of the environment that the policy represents has
become obsolete. If the model continues to be updated and a new policy
is recomputed every certain period, the resources (like time or capacity)
required for such computation in an online context should also be
considered. Also, if possible, it is necessary to determine the periodicity
with which to perform such updates. In this line, the approaches in~\citet{Barrett2011,Gari2019}
present this limitation because the policies are learned in \emph{offline}
mode.

\emph{Model-based and Model-free techniques:} \emph{\flqq{}Difficulty
to manage large state spaces}\frqq{}. This limitation generally affects
both Model-based and Model-free methods. In the first case, the computational
complexity of the algorithms like Value Iteration and Policy Iteration
is polynomial in the number of states and actions defined. From the
two analyzed works~\citet{Gari2019,Barrett2011} in the Model-based
category, it can be seen that the state space and actions are limited.
On the other hand, in the works based on Model-free methods~\citet{Peng2015,Xiao2017,Duggan2017,Soualhia2018,Dutreilh2011,Ghobaei-Arani2018,Dezhabad2018,BibalBenifa2018,Barrett2012,Horovitz2018,Wei2019},
the problem of managing many states and actions is associated not
only with the requirements to store the function $Q(s,a)$ but also
with the time and amount of data needed to update it. For example,
the use of many features (or dimensions) generates a combinatorial
explosion of states that is very difficult to handle. Besides, when
scalability is required in some of the defined variables, the number
of states increases considerably depending on the possible values
\LyXZeroWidthSpace \LyXZeroWidthSpace of those variables. In this
sense, since the problem of the dimension of the state space (also
known as dimensionality problem) is one of the main limitations of
RL, different alternatives have been studied to try to mitigate its
effects. An alternative to this problem~\citet{Peng2015,Xiao2017}
consists of the \emph{aggregation of states} by defining certain ranges
of values \LyXZeroWidthSpace \LyXZeroWidthSpace for the variables
that define them, grouping similar states. A second alternative is
to combine RL with \emph{function approximation}, a kind of generalization
which is an instance of supervised learning. An example is the use
of non-linear approximations of $Q(s,a)$, as in the proposals that
use deep neural networks~\citet{Liu2017,Cheng2018,Tong2020,Du2019}.
However, RL with function approximation has not yet been fully exploited
in the area of Cloud autoscaling. A third option is the use of fuzzy
logic combined with RL~\citet{T.Veni2016,Arabnejad2017}, for evolving
rules that enable approximate reasoning.

\emph{Model-free techniques:} \flqq{}\emph{Slow convergence}\frqq{}.
In the basic variants of Model-free methods, the function $Q(s,a)$
is updated when an action is executed, but only the visited state
value at that time is updated. Although convergence is guaranteed,
this usually involves a long training time, especially when it comes
to problems with many states and/or many actions. To reduce training
time there are proposals~\citet{Barrett2012,BibalBenifa2018,Nouri2019}
with multiple agents that learn in parallel and share the obtained
information. On the other hand, in~\citet{Dutreilh2011} the authors
propose to accelerate convergence using ideas from Dynamic Programming.
For this, frequent phases of updating the function $Q(s,a)$ are defined
using estimations of the obtained state values by recording the observations
made of the visited states, transitions, and rewards.

\emph{Model-free techniques:} \emph{\flqq{}Poor initial performance\frqq{}}.
Considering that at the beginning of the learning process there is
not an adequate policy (cold start effect), the initial performance
of the strategy is usually poor and it will be improved as it converges
to an appropriate policy. From all surveyed works, only in~\citet{Dutreilh2011}
we found a proposal to address this problem using an initial approximation
of $Q(s,a)$.

\subsection{Open Possibilities}

RL has demonstrated a great potential for automatically solving decision-making
problems, particularly because of their ability to consider long-term
consequences of the available actions. Some of the most impressive
results have been shown in Game Theory~\citet{Silver2016,Mnih2015},
but the potential of RL can be also extended to many other areas.
Specifically, in the case of Cloud autoscaling, only the first steps
have been taken, and much remains to be done. From the analysis of
the State of the Art, it becomes evident that there is a long list
of current limitations which in turn means that there is a wide spectrum
of research opportunities regarding RL techniques in the area of \LyXZeroWidthSpace \LyXZeroWidthSpace Cloud
autoscaling. In the general sense, there is a wide number of unexplored
combinations derived from the taxonomies outlined earlier in Section~\ref{subsec:AnalysisState-of-the-art}.

It would be interesting to design and develop autoscaling strategies
for scientific applications in Cloud that combine a scheduler and
scaler, both based on RL. These strategies could be based on the learning
of appropriate scheduling and scaling policies, which allow dealing
with the inherent uncertainty in the execution of applications in
the Cloud. Besides, when policies are learned in an online mode, they
would be able to adapt to changes in the dynamics of the environment.
In the context of Cloud application execution, the uncertainty comes
from the variability in the performance of the Cloud infrastructure,
and also, the changes in the environment may be due to adjustments
in the instances prices (as resource prices depend on market-like
fluctuations) and even due to the appearance of other types of instances
with different performance-cost trade-offs. The scaling policy could
try to adjust the infrastructure dynamically according to the variable
demand of the application, while the scheduling policy could determine
the most appropriate resource for the execution of each task, considering
the characteristics of each task and the available infrastructure.
Both policies would be learned from experience in the interaction
with the Cloud environment; modifying it and observing the effects.

Regarding the learning process, \emph{parallel learning} is a topic
that deserves much more attention. Parallel learning schemes update
the Q values in parallel, speeding up the process of policy learning.
In real Cloud settings, this kind of scheme might have special importance
since multiple autoscaling agents could share the feedback derived
from their actions and update the Q-values collectively. From a theoretical
point of view, this accelerates policy convergence but also allows
that an enormous amount of agents operate as feedback collectors while
at the same time are benefiting from the latest Q updates making the
information instantly available to all of the agents.

Nowadays, \emph{one-step}, \emph{tabular}, \emph{model-free} TD is
the most widely used RL methods. This is probably due to their great
simplicity, but these algorithms can be extended making them slightly
more complicated and significantly more powerful (i.e., multi-step
forms, various forms of function approximation rather than tables,
etc.)~\citet{Sutton:2018}. In the area of Cloud autoscaling, the
majority of approaches still use the simplest variants of RL methods.
Therefore, there is still room for further investigating the synergy
of different variants of the basic RL strategies and other machine
learning methods. In fact, 20\% of the surveyed works, which have
been published in 2017-2020, have exploited deep neural networks,
which shows a trend in this line.

Also, the particular states, actions, and how they are represented
can strongly affect the performance of the implemented approach. In
RL, as in other areas of Machine Learning, such representational choices
are, nowadays, more an art than a science~\citet{Sutton:2018}. In
the area of Cloud autoscaling, it is fundamental to study the specific
implications of such representational choices (states and actions)
and how they impact the performance of autoscalers. For example, interesting
questions to answer in this matter are: (1) What information from
a real Cloud environment is relevant to properly learn a policy? (b)
What could be an adequate representation of this information to accelerate
the learning process?

On the other side, the problem of Cloud autoscaling is closely related
to Multi-objective Optimization (MOO). The reader might have noticed
that in almost all surveyed papers, multiple optimization objectives
are present. Even more, conflicting objectives (such as economic cost
and execution time) are common in this context. Current proposals
usually combine these objectives in the reward function try to optimize
all of them at the same time. However, many other possibilities are
investigated in the active area of research called Multi-objective
Reinforcement Learning (MORL)~\citet{Liu2015}, which combines the
concepts and strengths of such two important fields: MOO and RL. Needless
to say that the study of MORL techniques in Cloud autoscaling is a
fundamental future line of research but yet it is incipient.

\section{Conclusions \label{sec:Conclusions}}

The flexibility and elasticity offered by the Cloud Computing paradigm
have opened opportunities to the study of autoscaling strategies for
the efficient execution of workflows, independent tasks, and Cloud
service applications. However, the variability in Cloud performance
generates an important uncertainty factor when making scaling and/or
scheduling decisions during application execution. In this sense,
RL-based strategies allow autoscalers to learn appropriate policies
through interaction with a stochastic environment. In this context,
recent research is focused on the exploitation of RL-based strategies
to address the autoscaling subproblems, i.e. scaling and/or scheduling.

Motivated by these facts, we have surveyed and classified works in
this area by deriving a taxonomy according to the type of RL-based
technique used. On the first level of the taxonomy, proposals in the
Model-based and the Model-free categories are presented. Then, on
a second level, proposals in the Model-free category are classified
into three groups. First, are those proposals that apply the technique
in its original or pure formulation. These techniques are further
subdivided into sequential or parallel, since the variant of RL isgiven
by parallel learning. Second, we present the proposals that combine
RL with neural networks, and finally, the proposals that combine RL
with Fuzzy Logic (FL).

As evidenced in the analysis of the reviewed literature, algorithms
based on RL such as \emph{Q-learning} and \emph{SARSA} have shown
to be effective in the online learning of scaling and scheduling policies
in the Cloud. A 45\% of the surveyed works are based on the autoscaling
problem in Cloud for workflow and independent tasks applications,
which are applications with distinctive features (long-term tasks,
data-intensive or computational-intensive tasks, high workload variability
with high parallelism and bottleneck stages) mostly used in engineering
and scientific settings, while a 55\% of the works focus on Cloud
service applications, mostly used in e-commerce and business settings.
However, a major finding is that neither of the surveyed works proposes
a solution that covers both the scaling and scheduling problems. Hence,
the inception of full-fledged autoscalers purely based on RL techniques
for either type of Cloud application remains to be seen in the area.

As a final comment, it is important to note that RL is a key technology
for the future development of Distributed Computing systems, even
beyond Cloud Computing. In particular, through RL it would be possible
to develop autonomous infrastructure management platforms that meet:
\begin{itemize}
\item Transparency: Implementation and operation details of the applications
would be hidden to the user. The use of these systems would not depend
on human intervention nor would demand to have access to deep domain
knowledge since it is expected that scaling and planning policies
are learned through the interaction with the environment. Such a scenario
would be different from the actual one in which, for example, the
scaling approaches that public computing infrastructures~\citet{AmazonAutoescaling}
use, are based on explicit thresholds for resource use. Such thresholds
are usually defined by experts based on the available metrics such
as CPU or memory usage.
\item Dynamism: At any moment, learned policies would allow the provider
to take the necessary actions given the current state of the environment
and the state of the applications. In such a scenario, the system
would not have to rely on static plans nor on rule-based actions defined
manually.
\item Adaptability: Thanks to online learning, policies can be constantly
improved and updated. In such a way, the policies would be able to
adapt to the changes that occur in the dynamics of the execution environment.
Such a characteristic is fundamental compared with policies learned
in offline mode~\citet{Gari2019} that are prone to become obsolete
in time.
\end{itemize}
Although the potential benefits are evident, many efforts are still
necessary towards making these goals a reality.

\section*{Acknowledgments}

We acknowledge the financial support by CONICET, grant number PIP
2017-2019 GI 11220170100490CO, and the SIIP-UNCuyo projects No. B082
and 06/B369.

\bibliographystyle{unsrt}
\addcontentsline{toc}{section}{\refname}\bibliography{itic,bibliography}

\subsection*{A.1 MDP Resolution via Dynamic Programming\label{subapp:DynamicProgramming}}

Dynamic programming (DP) in this context refers to a collection of
algorithms that can be used to compute optimal policies given a perfect
model of the environment as an MDP. Methods based on DP compute the
policy based on a complete model of the environment (Model-based).
In the process of estimating the state values $V(s)$, the probability
distribution of the transitions between the states $P_{a}(s,s')$
is used. This often becomes a limitation, since it is not always possible
to derive such a model. In some cases, the probability distribution
of the transitions is estimated from data obtained in previous experiences.
The DP methods offer an \emph{offline} learning variant, where the
policy is obtained by iterating over the model and not based on the
dynamics of current experiences. It is important to note that prior
estimates of other states are used in the process of estimating the
values \LyXZeroWidthSpace \LyXZeroWidthSpace of the states (a technique
known as \emph{bootstrap}). Two widely used DP algorithms are \emph{policyIteration}
and \emph{valueIteration}. Both algorithms have polynomial complexity
in the number of states and actions, so it is important to consider
the dimensions of these spaces when using DP. However, the search
performed with DP is much more efficient than an exhaustive exploration
in the space of all possible policies.

The \emph{policyIteration} algorithm\emph{ }(see Algorithm~\ref{alg:Policy-Iteration})
is defined based on the iterative repetition of the evaluation and
the improvement of the policy until convergence is achieved. In this
way, the algorithm generates the following sequence of value functions
and policies $v_{0}\rightarrow\pi_{0}\rightarrow v_{1}\rightarrow\pi_{1}...\rightarrow\pi^{*}$
until to reach an appropriate policy. On the other hand, the \emph{valueIteration}
algorithm (see Algorithm~\ref{alg:Value-Iteration}), first includes
the search for the appropriate value function and then, the computation
of the associated policy. These steps are not repeated because once
the value function is adequate, so is the associated policy. The search
for the appropriate value function can be understood as a combination
of the policy improvement process and a truncated evaluation of the
policy (the values \LyXZeroWidthSpace \LyXZeroWidthSpace are reassigned
after a single sweep of the states) without losing convergence~\citet{Sutton:2018}.
In this way, the algorithm generates the sequence of value function
updates $v_{0}\rightarrow v_{1}\rightarrow v_{2}\rightarrow...\rightarrow v^{*}$
and then, it computes the suitable policy $\pi^{*}$.

Both algorithms, \emph{policyIteration} and \emph{valueIteration},
formally require an infinite number of iterations to converge exactly
to the appropriate policy. In practice, both algorithms stop when
the difference between two successive approximations is less than
a limit $\Theta$, which usually within a much lesser number of iterations\citet{Sutton:2018}.

\begin{algorithm}[H]
\caption{\label{alg:Policy-Iteration}The Policy Iteration algorithm.}

\begin{algorithmic}[1]
\Procedure{PolicyIteration}{$S,A,P,R,\gamma,\Theta$}:
	\State{1.Initialize $V(s)$ y $\pi(s)$ arbitrarily $\forall s \in S$}
	\State{2.Policy Evaluation}
	\Repeat:
	\State {$\Delta \leftarrow 0$}
	\ForEach{$s \in S$}
		\State {$v \leftarrow V(s)$}
		\State {$a \leftarrow \pi(s)$}
		\State {$V(s) \leftarrow \sum_{s'}P_{a}(s,s')[R_{a}(s,s')+\gamma V(s')]$}
		\State {$\Delta \leftarrow \max(\Delta,\abs{v-V(s)})$}
	\EndFor
	\Until {$\Delta < \Theta$ (a small positive number)}
	\State {3.Policy Improvement}
	\State {$stablePolicy \leftarrow true$}
	\ForEach{$s \in S$}
		\State {$oldAction \leftarrow \pi(s)$}
		\State{$\pi(s)\leftarrow\arg\max_{a}\sum_{s'}P_{a}(s,s')[R_{a}(s,s')+\gamma V(s')]$}
		\State {{\bf If}  $oldAction \neq \pi(s)$ {\bf then} $stablePolicy \leftarrow false$ }
	\EndFor
\State {{\bf If}  $stablePolicy$ {\bf then} stop and return $V \approx v*$ and $\pi \approx \pi*$ }
\EndProcedure  
\end{algorithmic}
\end{algorithm}

\begin{algorithm}[H]
\caption{\label{alg:Value-Iteration}The Value Iteration algorithm.}

\begin{algorithmic}[1]
\Procedure{ValueIteration}{$S,A,P,R,\gamma,\Theta$}:
	\State{Initialize array $V$ arbitrarily $\forall s \in S$} 
	\Repeat:
	\State {$\Delta \leftarrow 0$}
	\ForEach{$s \in S$}
		\State {$v \leftarrow V(s)$}
		\State {$V(s) \leftarrow \max_{a}\sum_{s'}P_{a}(s,s')[R_{a}(s,s')+\gamma V(s')]$}
		\State {$\Delta \leftarrow \max(\Delta,\abs{v-V(s)})$}
	\EndFor
	\Until {$\Delta < \Theta$ (a small positive number)}
	\State {Output a deterministic policy $\pi \approx \pi*$  such that:}
\State{$\pi(s)\leftarrow\arg\max_{a}\sum_{s'}P_{a}(s,s')[R_{a}(s,s')+\gamma V(s')]$}
  \EndProcedure  
\end{algorithmic}
\end{algorithm}

\subsection*{A.2 MDP Resolution via Temporal Difference\label{subapp:TemporalDifference}}

Methods based on Temporal Difference (TD) do not require a perfect
model of the environment (Model-free), since the policy learning process
is based on the observed dynamics during its experimentation. In this
sense, these methods offer an approach with a greater ability to adapt
to changes in the environment, since unlike DP-based methods, learning
through TD occurs in an \emph{online} way. Because of the lack of
the model, the state value function $V(s)$ is not sufficient in suggesting
a policy and it is required to estimate the values related to each
action. The action-value function $Q(s,a)$ represents the expected
gain considering the state-action pair and it is usually represented
in tabular form. Similarly to DP, to estimate new values, the previously
estimated values \LyXZeroWidthSpace \LyXZeroWidthSpace are used (i.e.,
bootstrap is performed).

Two widely used algorithms in this area are \emph{Q-learning}~\citet{Watkins1992}
and \emph{State-Action-Reward-State-Action} (\emph{SARSA}). It is
important to highlight that one of the main limitations in RL is that
the convergence time of these algorithms depends directly on the dimension
of the state space and actions. Moreover, since these algorithms do
not have an adequate initial policy they have a poor initial performance
that will have a greater or lesser impact depending on the addressed
problem and the time taken for training. Making inappropriate decisions
at the beginning of the autoscaling of workflows in Cloud, which is
necessary for the exploration process, can have a direct impact on
the makespan and the economic cost, so it is convenient to have an
acceptable initial policy. This could also reduce the time required
to learn the right policy. In any case, it is necessary to consider
that obtaining an acceptable initial policy is not always trivial~\citet{Dann2014}.

The distinctive characteristic of \emph{Q-learning} (see Algorithm~\ref{alg:Q-learning})
is that it uses two different policies, one to select the next action
and another to update $Q$. In other words, \emph{Q-learning} tries
to evaluate $\pi$ while following another policy $\mu$. Alternatively,
SARSA (see Algorithm~\ref{alg:SARSA}) uses the same policy all the
time. The most important difference between the two above mentioned
algorithms is how $Q$ is updated after each action. \emph{Q-learning}
updates $Q$ with the action that maximizes the gain for the next
step. This makes \emph{Q-learning} follows an \emph{$\varepsilon$-greedy
}policy\emph{}\footnote{\emph{$\varepsilon$-greedy}: a policy that with an $\varepsilon$
probability selects a random action, but most of the time it selects
an action with the maximum estimated value.} with $\varepsilon=0$, i.e., there is no exploration. In contrast,
SARSA updates $Q$ by following exactly an \emph{$\varepsilon$-greedy}
policy, since the action is extracted from it. Both algorithms include
the $\alpha\in(0,1]$ parameter relative to the size of the step in
the learning process, and the $\varepsilon>0$ parameter that determines
the exploration degree of new policies.

\begin{algorithm}[H]
\caption{\label{alg:Q-learning}The Q-learning algorithm.}

\begin{algorithmic}[1]
\Procedure{Q-learning}{$S,A,P,R,\gamma,\alpha,\varepsilon$}:
	\State{Initialize $Q(s,a)$ arbitrarily $\forall s \in S, a \in A$ y $Q(terminalState,.)=0$} 
	\ForEach{(episode)}
	\State{Initialize S}
	\Repeat:
		\State {Select A from S using the policy derived from de Q ($\varepsilon$-greedy)}
		\State {Take action A, observe R,S'}
		\State {$Q(S,A) \leftarrow Q(S,A) + \alpha[R+\gamma \max_a Q(S',a)-Q(S,A)]$}
		\State {$S \leftarrow S'$}
	\Until {S is terminal}
	\EndFor
  \EndProcedure  
\end{algorithmic}
\end{algorithm}

\begin{algorithm}[H]
\caption{\label{alg:SARSA}The SARSA algorithm.}

\begin{algorithmic}[1]
\Procedure{SARSA}{$S,A,P,R,\gamma,\alpha,\varepsilon$}:
	\State{Initialize $Q(s,a)$ arbitrarily $\forall s \in S, a \in A$ y $Q(terminalState,.)=0$} 
	\ForEach{episode}
	\State{Initialize S}
	\State {Select A from S using the policy derived from Q ($\varepsilon$-greedy)}
	\Repeat:
		\State {Take action A, observe R,S'}
		\State {Select A' from S' using the policy derived from Q ($\varepsilon$-greedy)}
		\State {$Q(S,A) \leftarrow Q(S,A) + \alpha[R+\gamma Q(S',A')-Q(S,A)]$}
		\State {$S \leftarrow S'$; $A \leftarrow A'$}
	\Until {S is terminal}
	\EndFor
  \EndProcedure  
\end{algorithmic}
\end{algorithm}

\subsection*{A.3 MDP Resolution via Neural Networks\label{subapp:Approach-DeepRL}}

Large state spaces in an RL problem leads to the need to find non-tabular
representations of the $Q$ function, not only for the memory required
to store large tables, but also for the time it would take to fill
it. Algorithms capable of generalizing in more complex and sophisticated
state space contexts are then consequently needed. In this sense,
non-linear approximations of $Q$ with artificial neural networks
appeared. A type of neural network that has proven very successful
in RL applications~\citet{Mnih2015,Silver2016} is Deep Convolutional
Neural Networks, which are specialized in the processing of large-scale
data organized in spatial matrices. Then, the strategy that combines
RL with Deep Neural Networks (DNN) is called Deep Reinforcement Learning
(DRL). Finally, the DNN used to approximate the $Q$ function is called
\emph{DeepQNetworks} (DQN) and the learning algorithm that uses DQN
is referred to as \emph{Deep Q-learning}.

\begin{figure}[H]
\begin{centering}
\includegraphics[scale=0.8]{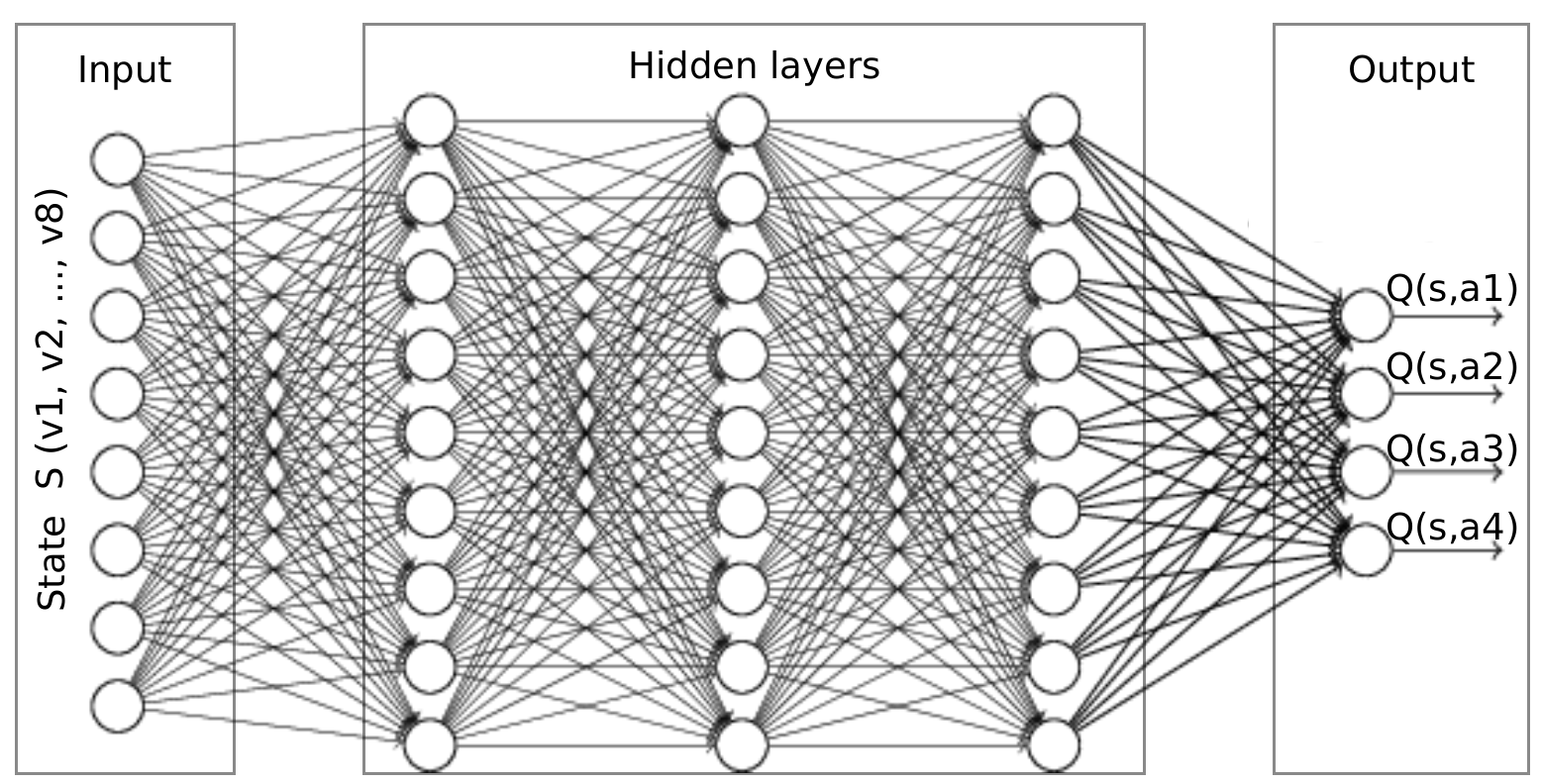}
\par\end{centering}
\caption{\label{fig:DeepQNetwork}Example of the structure of a DQN.}
\end{figure}

Figure~\ref{fig:DeepQNetwork} shows an example of a DQN. The input
corresponds to a state $s$ of the environment and the output represents
the estimated value of function $Q$ for the state $s$ and all possible
actions. In the process of training the network, the objective is
to minimize the approximation error between the result of the network
and the optimality equation of Bellman~\citet{Mnih2015}. Thereby,
the same problem as with the classical DP and TD techniques is solved,
but now using a nonlinear approach based on deep neural networks.

\subsection*{A.4 MDP Resolution via Fuzzy Logic\label{subapp:Approach-FL}}

Fuzzy Logic (FL) appears as another alternative to address the dimensionality
problem of RL strategies. The idea is to reduce the state space using
a diffuse representation of the information.

\begin{figure}[H]
\begin{centering}
\includegraphics[scale=0.8]{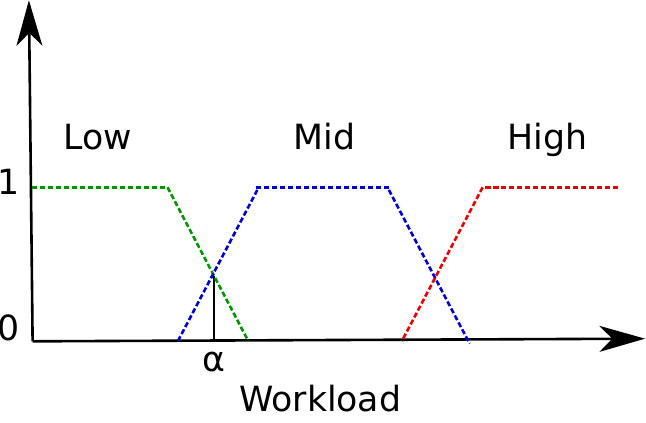}
\par\end{centering}
\caption{\label{fig:workload}Example of the fuzzy membership function (Y axis)
for the workload variable. The fuzzy sets are defined using a trapezoid.}
\end{figure}

Broadly, FL systems attempt to represent knowledge inaccurately, similar
to how human beings do, and as opposed to classical numerical forms.
In this sense, FL works with fuzzy sets in which the elements have
some membership degree. To define the membership function of these
sets, triangles or trapezoid curves are usually used (see Figure~\ref{fig:workload}).
For example, in Figure~\ref{fig:workload}, a fuzzy membership function
is represented for a \emph{Cloud workload} variable with three fuzzy
sets (Low, Medium, High) that define the membership degree of the
variable to each of them. Thus, in the presence of a workload $\alpha$,
it is possible to affirm that it belongs both to the Low and Medium
fuzzy sets with a 50\% of probability, and hence the diffuse nature
of this representation.

These concepts from FL allow reasoning based on rules of the form:

\[
\mathrm{if}(antecedent)\mathrm{then}(consequent),
\]
where the antecedent and consequent values are expressed in a fuzzy
way. Based on the previous example, one possible rule is: if the workload
is \emph{high} then \emph{more VMs} must be allocated.

FL has been applied in different fields, from Control Theory to Artificial
Intelligence. A control process based on FL consists of the following
steps:
\begin{itemize}
\item Mapping of input data to fuzzy set labels (Fuzzifier)
\item The inference process based on fuzzy rules (Fuzzy Reasoning)
\item Fuzzy output mapping to clear values (Defuzzifier)
\end{itemize}
Fuzzy Reinforcement Learning (FRL) is the strategy that combines the
strength of fuzzy reasoning with RL. FRL allows handling problems
with large state spaces without affecting the performance of the RL
algorithms. For this, in the learning process a fuzzy representation
of the information is used, which considerably reduces the number
of states.
\begin{figure}[tbh]
\begin{centering}
\includegraphics[scale=0.5]{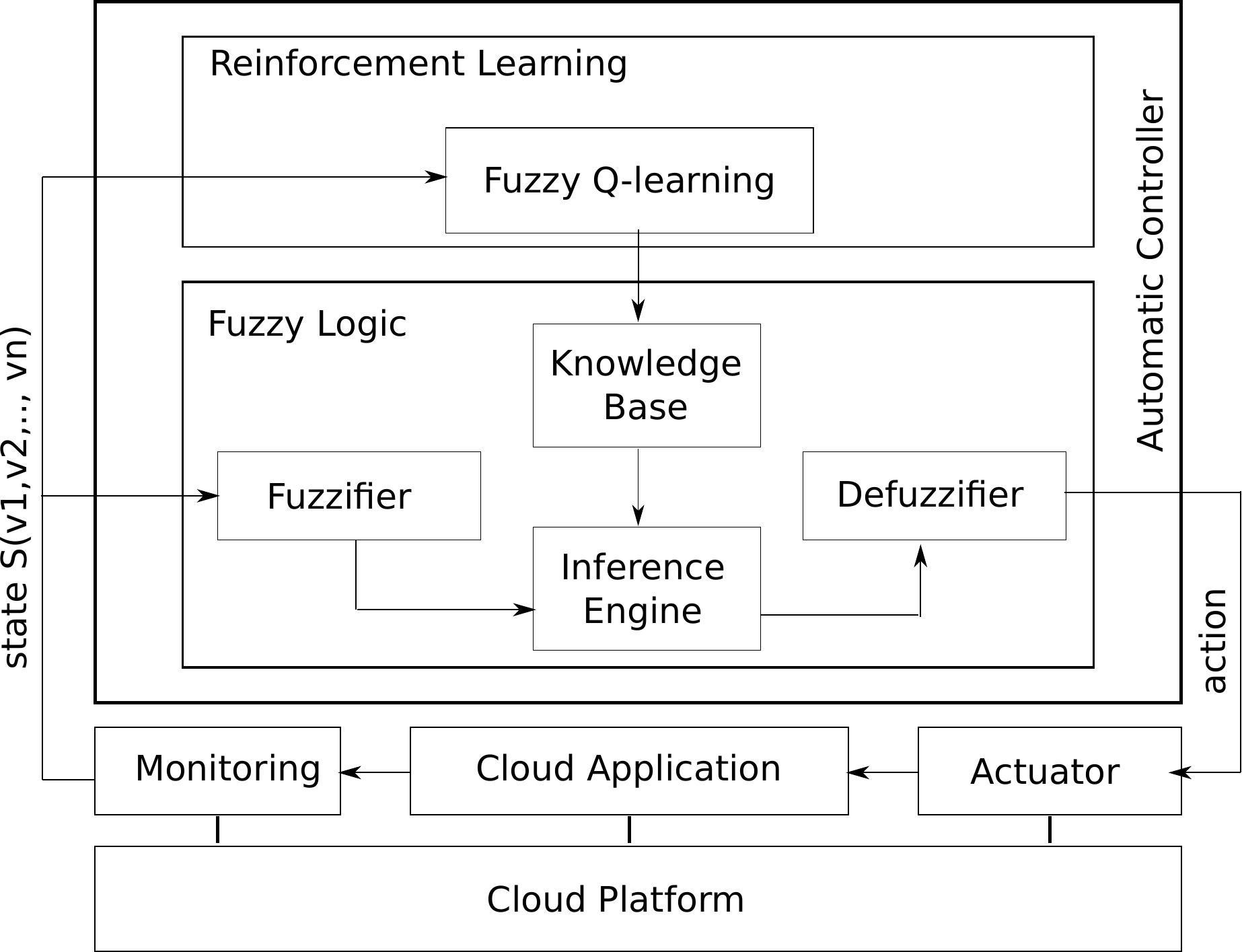}
\par\end{centering}
\caption{\label{fig:RL-LD}Example of the architecture of a Cloud autoscaling
system based on RL and FL. Figure adapted from~\citet{Jamshidi2016}.}
\end{figure}

Motivated by this benefit, some authors~\citet{Arabnejad2017,T.Veni2016}
have proposed approaches based on FRL for autoscaling in Cloud. Figure~\ref{fig:RL-LD}
shows the interaction between the components involved in these approaches.
First, the Cloud platform and the running application that composes
the environment, which is continuously observed by a monitoring process.
The monitoring process retrieves data of interest in the state of
the environment and reports it to the Automatic Controller (AC). One
of the main components of the AC is the FL-based control process called
Fuzzy Controller (FC). The FC is composed of the Knowledge Base (or
rules), the Fuzzifier, the Inference Engine and the Defuzzifier. In
this way, the FC receives the signal of the environment state, transforms
it to its diffuse representation, reasons based on the rules, and
obtains a diffuse output that is finally returned to its clear representation.
This output or action is used by the \emph{actuator process} to modify
the environment. The second component of the FC is precisely the RL
process, which also receives the signal of the environment state and,
guided by the optimization objectives, is responsible for learning
the most appropriate set of rules to update the knowledge base of
FC. Each member of the table of values $Q$ \LyXZeroWidthSpace \LyXZeroWidthSpace is
assigned to a specific rule (which describes some action-state pairs).
Then, these values \LyXZeroWidthSpace \LyXZeroWidthSpace are updated
during the learning process. In this way, it is possible to take advantage
of the strengths of RL and FL strategies to design an automatic controller
capable of evolving fuzzy rules that allow making approximate reasoning.
\end{document}